\documentclass{aa} 

\usepackage{graphicx} 
\usepackage{txfonts} 
\usepackage{natbib} 
 \usepackage{rotating,amsmath,amsfonts,amssymb}
\usepackage{hyperref}

\newcommand{\Mpc}{$h^{-1}$\thinspace Mpc}

\newcommand{\be}{\begin{equation}}
\newcommand{\ee}{\end{equation}}
\newcommand{\bea}{\begin{equation}\begin{aligned}} 
\newcommand{\eea}{\end{aligned}\end{equation}}

\newcommand{\eg}{\emph{e.g.}}
\newcommand{\dthree}{\delta_{\rm 3D}}
\newcommand{\dtwo}{\delta_{\rm 2D}}
\newcommand{\xithree}{\xi_{\rm 3D}}
\newcommand{\xitwo}{\xi_{\rm 2D}}
\newcommand{\per}[1]{{\bf #1}_\perp}
\newcommand{\parl}[1]{#1_\parallel}
\newcommand{\dd}[1]{{\rm d}#1\,}

\def\apj{ApJ} 
\def\apjl{ApJL} 
\def\apjs{ApJS} 
 
\def\aap{A\&A} 
\def\mnras{MNRAS}

\usepackage{color}

\begin{document}    
 
\title{Correlation functions in 2D and 3D as descriptors of the cosmic
web}
 
\author{J. Einasto\inst{1,2,3} 
\and  G. H\"utsi\inst{4} 
%
\and   M. Einasto\inst{1}  
}
 
\institute{Tartu Observatory, 61602 T\~oravere, Estonia 
\and 
Estonian Academy of Sciences, 10130 Tallinn, Estonia
\and  
ICRANet, Piazza della Repubblica 10, 65122 Pescara, Italy 
\and
National Institute of Chemical Physics and Biophysics, 
Tallinn 10143, Estonia } 
 
\date{ Received 07 April 2020; accepted 25 May 2021}  
 
\authorrunning{Einasto et al.} 
 
\titlerunning{Correlation functions and the cosmic web} 
 
\offprints{J. Einasto, e-mail: jaan.einasto@ut.ee}

\abstract {} 
{Our goal is to find the relation between the two-point correlation
  functions (CFs) of projected and spatial density fields of
  galaxies in the context of the cosmic web. }
 {To investigate relations between spatial (3D) and projected (2D) CFs
   of galaxies we used density fields of two simulations: a
   $\Lambda$-dominated cold dark matter (LCDM) model with known
   particle data, 
   and the Millennium simulation with know data on simulated
   galaxies. We compare 3D and 2D correlation functions. In the 2D case, we
   use samples of various thickness to find the dependence of 2D CFs on
   the thickness of samples. We also compare 3D CFs in real and redshift
   space.  }
{The dominant elements of the cosmic web are clusters and filaments,
  separated by voids filling most of the volume.  In individual 2D
  sheets, the  positions of clusters and filaments do not coincide. As a
  result, in projection, the clusters and filaments fill in 2D voids. This
  leads to a decrease in the amplitudes of CFs (and power spectra) in
  projection.  For this reason, the amplitudes of 2D CFs are lower than
  the amplitudes of 3D correlation functions: the thicker the 2D sample, the greater the difference.  }
{Spatial CFs of galaxies contain valuable information about the geometrical
  properties of the cosmic web that cannot be found from projected CFs. }
\keywords {Cosmology:
  large-scale structure of the Universe; Cosmology: dark matter;
  Cosmology: theory; Galaxies: clusters; Methods: numerical}

\maketitle

\section{Introduction}

{The early Universe was almost smooth, with only tiny
  fluctuations generated during the early inflationary stages of
  evolution.  During the early stages, the fluctuations evolved
  independently, like linear waves on the surface of a body of deep water 
  \citep{Coles:2000aa}.  Overdense fluctuations attract additional
  mass as the Universe expands \citep{Peebles:1980aa}.  As the
  structures grew in mass, they interacted with others  in a non-linear
  way, like waves breaking in shallow water \citep{Coles:2000aa}, and
  formed the presently observed cosmic web.  

  Density fields possess Fourier modes with real and imaginary parts,
  which are independently distributed.  Most commonly used statistical
  methods to describe the structure of the Universe are power spectra
  and correlation functions (CFs).  These describe real parts of the
  density distribution.  However, to describe the pattern of the
  cosmic web, the phase information is needed.  This was demonstrated
  in Fig.~1 of \citet{Coles:2000aa}, where the left panel shows a slice
  of the
  density field found by a numerical simulation of the evolution of
  the Universe, and the right panel shows a slice of the same field,
  generated by randomly reshuffling the phases of Fourier modes.  In
  the right panel, the whole structure of the web is gone. Both fields
  have identical power spectra, but completely different information
  content. However,  the CF contains information which in a specific way
  describes some properties of the pattern of the density field. }

{ Early data allowed projected  two-dimensional (2D) CFs to be found from
observational data.  In the 1980s, redshift data became available and it
was possible to find spatial three-dimensional (3D) CFs. It is known
that redshifts are distorted by local movements of galaxies in
clusters -- the finger of god (FOG) effect, and that galaxies
and clusters flow toward attractors -- the \citet{Kaiser:1987aa} effect.
These effects influence 2D and 3D CFs differently.  Also, the spatial
structure of the cosmic web, as described by its fractal character, is
different in 2D and 3D density fields and respective CFs
\citep{Einasto:2020aa}. 

The goal of this paper is to study the relationship between projected
2D and spatial 3D density fields, and 2D and 3D CFs in the context of
the cosmic web.  It is well known that observational samples are
influenced by selection and border effects.  To avoid complications
caused by these effects, we  study the relationship between
projected and spatial density fields and respective CFs using
simulated dark matter (DM) models.  We assume that the evolution and
the present structure of the Universe can be well described by the
$\Lambda$-dominated cold dark matter (LCDM) model.}

To calculate CFs, we use a novel method developed by
\citet{Szapudi:2005aa}.  This method uses density fields on 3D or 2D
grids as input data, and applies fast Fourier transform (FFT) to
calculate CFs, in a similar way to calculations of power spectra.  In
this way, we are able to look at CFs from a different point of view,
as descriptors of the continuous density field of the cosmic web with
its complex pattern. Some aspects of CFs as descriptors of the cosmic
web were studied by \citet{Einasto:2020aa}; here we continue this
discussion.

To find the relationship between projected and spatial density fields
and CFs, we use two LCDM models. One model was
calculated by our team in a box of size 512~\Mpc.  We take advantage
of the fact that for this model, the positions of all particles are known,
and we can use these particles as objects in the cosmic web to study
the properties of the web.  The other model we use is the Millennium
simulation, which has a box of size 500~\Mpc.  For our study,
we use the galaxy catalogue based on a semi-analytical model of galaxy
formation.  For the correlation analysis, we use simulated galaxies as
test particles.

The paper is organised as follows.  In the following section we describe
our simulation data, the methods to calculate density fields, the  CFs, and their
derivatives. In section 3, we compare spatial and projected CFs and
their dependence on input parameters -- particle density (luminosity),
threshold of model samples, and on the thickness of projected shells
of the 2D density field.  In section 4 we compare the
properties of spatial and projected density fields and CFs as
descriptors of the cosmic web.  We present our conclusions in the 
final section.

\section{Data and methods}

In this section we describe our LCDM model and the galaxy
catalogue of the Millennium simulation.  We also describe methods to
calculate 2D and 3D CFs.

\subsection{ LCDM   model samples}

We performed a simulation of the evolution of the cosmic web in a box
of size $L_0=512$~\Mpc, with resolution $N_{\mathrm{grid}} = 512$ and
with $N_{\mathrm{part}} = N_{\mathrm{grid}}^3$ particles. The initial
density fluctuation spectrum was generated using the COSMICS code by
\citet{Bertschinger:1995}, assuming $\Omega_{\mathrm{m}} = 0.28$,
$\Omega_{\Lambda} = 0.72$, close to concordance LCDM cosmological
parameters \citep{Bahcall:1999aa}, $\sigma_8 = 0.84$, and the
dimensionless Hubble constant $h = 0.73$.  To generate initial data we
used the baryonic matter density $\Omega_{\mathrm{b}}= 0.044$.
Calculations were performed with the GADGET-2 code by
\citet{Springel:2005}. The same model was used by
\citet{Einasto:2019aa} to investigate the biasing phenomenon, and by
\citet{Einasto:2020aa} to study general biasing and fractal properties
of the cosmic web. The model was described in earlier papers; for
consistency, we provide the basic data of the model below.

{\scriptsize 
\begin{table*}[ht] 
  \caption{Parameters of LCDM particle-density-limited samples}
\label{Tab1}                         
\centering
\begin{tabular}{lrlllc}
\hline  \hline
Sample   & $\rho_0$&$F_C$ &$FF_C$& $A$ & $\gamma$\\  
\hline  
(1)&(2)&(3)&(4)&(5)&(6)\\ 
\hline  
LCDM.00   &  0 & 1.000 &1.0000 & 0.729&   $-1.852  \pm   0.023$\\
LCDM.01   &  1 & 0.797  &0.3434&  1.203&  $-1.835  \pm   0.022$\\
LCDM.02   &  2 & 0.678  &0.2159& 1.530&  $-1.890  \pm   0.021$\\
LCDM.05   &  5 &  0.516 &0.10743 & 2.061& $-1.982  \pm   0.024$ \\
LCDM.10 &  10 & 0.4036  &0.05972 & 2.507& $-2.070  \pm   0.030$ \\
LCDM.20 &  20 & 0.3011 &0.03146 & 3.021&  $-2.165 \pm    0.040$ \\
LCDM.50 &  50 & 0.1831  &0.01169 &3.725& $-2.315  \pm   0.058$ \\
\hline 
\end{tabular} 
\tablefoot{
The columns are:
(1) sample name; 
(2)  particle-density limit $\rho_0$;
(3) the fraction of particles in the sample,  $F_{C}$; 
(4) total filling factor of all clusters at density threshold
$D_t=0.1$, $FF_C$;
(5) CF amplitude $A=\xi(6)$ at $r=6.0$~\Mpc;
(6) CF slope $\gamma$.} 
\end{table*} 
}

For all simulation particles, we calculated local density values at
particle locations, $\rho$, using the positions of the 27 nearest particles,
including the particle itself.  Densities were expressed in units of
the mean density of the whole simulation.  In the present study, we use
particle-density-selected samples at the present epoch. Model samples
contain particles above a certain limit, $\rho \ge \rho_0$, in units
of the mean density of the simulation.  Samples of particles of LCDM
models with particle density threshold $\rho \ge \rho_0$ are
considered as model equivalents of samples of galaxies of various
luminosity; see \citet{Einasto:2019aa}.  This particle-selection
algorithm is somewhat analogous to the Ising model of statistical
mechanics, which was implemented in cosmological studies by \citet{Repp:2019aa,
  Repp:2019ab}.  For the analysis, we used particle samples as given in
Table~\ref{Tab1}.  Particle-density-selected samples are referred to as
LCDM.$i$, where $i$ denotes the particle-density limit $\rho_0$.  The
full DM model includes all particles and corresponds to the
particle-density limit $\rho_0 = 0$, and is therefore denoted
LCDM.00.

{ The main parameters of the LCDM  model samples are given in
  Table~\ref{Tab1}.  We also provide the fraction of particles in the
  sample, $F_{C}= N_C/N_{\mathrm{part}}$, which is equal to the number
  density of selected particles, $\rho \ge \rho_0$, per cubic \Mpc.
  Next we give  the total filling factor, $FF_C$, of all
  particles with $\rho \ge \rho_0$, forming clusters in terms of the
  percolation theory \citep{Stauffer:1979aa} at density threshold
  $D_t=0.1$ for the density field of the LCDM model. This shows the
  fraction of the volume filled with selected particles.  Thereafter, we
  give the amplitude of the CF at $r=6$~\Mpc, $A=\xi(6)$; see
  below for definition.  Finally, we provide the slope of the 3D CF $\gamma$.}

{The local density value of particles, $\rho$, is an important
  parameter, and changes during the evolution.  The evolution of
  $\rho$ in our LCDM model was investigated by \citet{Einasto:2020ac},
  and in the Millennium simulation by \citet{Pandey:2013aa}.  Both
  authors found the density distribution of DM particles.  The
  fraction of particles, $F_C$, is an integral over the number of
  particles of various densities,
  $F_C(\rho_0)=\int_{\rho_0}^\infty
  N(\rho)\dd{\rho}/N_{\mathrm{part}}$. By construction it is a
  monotonically decreasing function. The total filling factor is also
  an integrated parameter, which describes the volume fraction of
  clustered particles.  }

\subsection{Millennium simulation galaxy samples}

We used the simulated galaxy catalogue by \citet{Croton:2006aa}, which
was calculated using a semi-analytical model of galaxy formation from
the Millennium simulations by \citet{Springel:2005aa}.  The Millennium
simulation was made in a cube of size 500~\Mpc\ using $2160^3$
particles, starting from redshift $z=127$.  The Millennium catalogue
contains 8\,964\,936 
simulated galaxies, with absolute magnitudes in {\rm r} colour
$M_r \le -17.4$, which correspond to galaxies brighter than the Small
Magellanic Cloud \citep{Springel:2005aa}.  The magnitudes are in
standard SDSS filters.  We extracted for our
analysis $x,y,z$ coordinates in \Mpc, velocities $v_x,v_y,v_z$ in
km/s, and absolute magnitudes in $r$ and $g$ photometric systems.  The
Millennium samples with $M_r $ luminosity lower limits of
$-17.4,~-18.0,~-19.0,~-20.0,~-20.5,~-21.0$, and $~-22.0$, are referred
to as Mill.17.4, Mill.18.0, Mill.19.0, Mill.20.0, Mill.20.5,
Mill.21.0, and Mill.22.0. We use luminosity threshold samples
with an upper magnitude limit of $-25.0$.  Parameters of the Millennium
luminosity-limited samples are given in Table~\ref{Tab2}.

{\scriptsize 
\begin{table*}[ht] 
\caption{Parameters of Millennium luminosity-limited galaxy samples} 
\label{Tab2}                         
\centering
\begin{tabular}{lcrcccc}
\hline  \hline
Sample   & $M_r $  & $N_{\mathrm{gal}}$ &
           $A_r$ &  $A_s$& $C$&$\gamma$\\                                                           
\hline  
(1)&(2)&(3)&(4)&(5)&(6)&(7)\\ 
\hline  
Mill.17.4 & $-17.4$ &  8\,964\,936&  0.922&1.191&  1.1368& $-1.816  \pm   0.010$ \\
Mill.18.0 & $-18.0$ &  6\,617\,818& 0.931 &1.202&  1.1359& $-1.809  \pm   0.009$ \\
Mill.19.0 & $-19.0$ & 3\,767\,977 & 0.929 &1.199&  1.1359& $-1.787  \pm   0.008$ \\
Mill.20.0 & $-20.0$ & 1\,882\,813  & 0.936&1.202&  1.1332& $-1.745  \pm   0.007$ \\
Mill.20.5 & $-20.5$ & 1\,183\,829  & 0.963&1.232&  1.1311&  $-1.721  \pm   0.007$\\
Mill.21.0 & $-21.0$ &  672\,909     & 1.086&1.375&   1.1249& $-1.718  \pm   0.009$ \\
Mill.22.0 & $-22.0$ &  104\,204     & 2.601&3.178&   1.1053&  $-1.904  \pm   0.030$ \\
\hline 
\end{tabular} 
 \tablefoot{
 The columns show 
(1) the sample name; 
(2) the absolute $r-$magnitude limit, $M_r $;
(3) the number of galaxies in the sample;
(4)  the CF amplitude  in
real space, $A_r=\xi(6)_r$;
(5)  the CF amplitude  in
redshift space, $A_s=\xi(6)_s$;
(6) the relative amplitude correction factor $C= \sqrt{\xi(6)_s/\xi(6)_r}$ of
CF in redshift space;
(7) CF slope $\gamma$.
}
\end{table*} 
} 

We perform the correlation analysis of Millennium galaxy samples
using galaxy positions in real space, as given by simulations, and in
redshift space.  In the second case, we assume that the observer is
located far away in the $z$-direction.  In this approximation, in
the calculation of 2D CFs the $x,y$ positions of galaxies are unaffected, but shifts are added
to $z$ positions corresponding to the galaxy
velocity $v_z$.  For both variants, we calculated the amplitude of the
CF at $r=6$~\Mpc, $A=\xi(6)$.  Amplitudes in real space,
$A_r=\xi(6)_r$, and in redshift space, $A_s=\xi(6)_s$, are given in Table
\ref{Tab2}. The last column  in Table~\ref{Tab2} is the slope of the real
space 3D CF $\gamma$.

\subsection{Calculation of spatial  correlation functions}

 Conventional methods cannot
be used to find the CFs of LCDM and Millennium samples because the
number of particles (galaxies) is too large.  To 
find CFs, we used  the \citet{Szapudi:2005aa} method.  This method
applies FFT to calculate CFs and scales as
$O(N \log N)$. {The method is an
implementation of the algorithm {\tt eSpICE}, the Euclidean version of
{\tt SpICE} by \citet{Szapudi:2001aa}.  The idea of the fast algorithm
is described by \citet{Moore:2001aa}. }
As input the method uses density fields on grids
$N_{\mathrm{grid}}^3$ for both models.  In calculations of the density fields we used
particle (galaxy) coordinates to put them into the grid of the density
field.  Particle local density values and galaxy magnitudes were used
as labels to select particles and galaxies for subsamples.  {For each
grid cell, we calculated the 3D density, $\rho({\bf x}) =N({\bf
  x})/\bar{N}$, and the density  contrast 
\be
\delta({\bf x}) = \rho({\bf x}) -1=   (N({\bf x})-\bar{N})/\bar{N};
\label{delta}
\ee
  where $N({\bf x})$ is the number of particles (galaxies) in the
cell at location $\bf x$, and
$\bar{N}$ is the mean number of particles/galaxies in a cell.}
The coordinates of all particles/galaxies are known, and therefore it is easy to
find density fields with higher resolution to resolve the structure on
small scales.

We tried several grid sizes from $N_{\mathrm{grid}}=1024$ to
$N_{\mathrm{grid}}=3078$.  These tests showed that on large scales
(particle separations) all grids give almost identical results.
This result has a simple explanation: for the CF, the fraction of
cells of different density is important, and this fraction remains
almost constant for different grid sizes.  On smaller scales, finer
grids allow us to investigate the internal structure of halos (clusters
of galaxies).  For the present analysis, the grid size
$N_{\mathrm{grid}}=2048$ is optimal. {The cell size of the density
  field is in this case 0.25~\Mpc, which is sufficient to test the
  internal structure of DM halos and simulated clusters of galaxies.
}

We calculated CFs for particle/galaxy separations up to
$L_{max}=200$~\Mpc\ with 90 logarithmical bins.  As an argument in
spatial 3D CFs, we used the pair separation in true 3D space,
$r=\sqrt{(\Delta\,x)^2+(\Delta\,y)^2+(\Delta\,z)^2}$. Here,
$\Delta\,x = x_i-x_j$, $\Delta\,y=y_i-y_j$, and $\Delta\,z=z_i-z_j$,
and $i$ and $j$ are particle (galaxy) indexes to compare.  {In
  classical studies \citep{Peebles:1975, Peebles:1980aa},  CFs were 
  characterised by the  mean slope in log-log presentation, $\gamma$,
  and by the correlation length, $r_0$, where the CF has a unit value of
  $\xi(r_0)=1$.  The present study shows that the information content
  of the CF is very rich. To describe various aspects of information
  content of the CF we  use  the slope of the CF as a
  parameter, $\gamma$, and  the log-log gradient of the pair correlation
  function, $g(r) = 1+\xi(r)$:
\begin{equation}
  \gamma(r)= {\dd{\log g(r)} / \dd{\log r}}.
  \label{gamma}
\end{equation}
 The gradient function $\gamma(r)$ characterises the shape of the
  CF.  On small scales it tests the internal structure of DM halos and
  clusters of galaxies; on larger scales it gives the fractal properties of the
  cosmic web \citep{Einasto:2020aa}. 
}

{Following \citet{Norberg:2001aa}, \citet{Tegmark:2004aa}, and
  \citet{Zehavi:2011aa} we derived   relative CFs,
  $\xi_C(r,\rho_0)/\xi_r(r)$,  for all models, which define the relative bias function:
\begin{equation}
  b_R(r, \rho_0) = \sqrt{\xi_C(r,\rho_0)/\xi_r(r)}, 
\label{biasCF}  
\end{equation}
where $\xi_C(r,\rho_0)$ is the CF of the clustered matter (particles in DM
halos or galaxies),  $\xi_r(r)$ is the CF of the reference sample, $r$ is the
separation of particles/galaxies, and $\rho_0$ is the particle density
limit, which selects particles to be included.  In Millennium samples,
instead of using $\rho_0$  as the argument, we use the magnitude limit of the
sample, $M_r$.  For our LCDM.00 model, we have the CF of all matter.
We use the CF of this sample as a reference CF to get the bias function
of clustered matter relative to all matter. }

\begin{figure*}[ht]
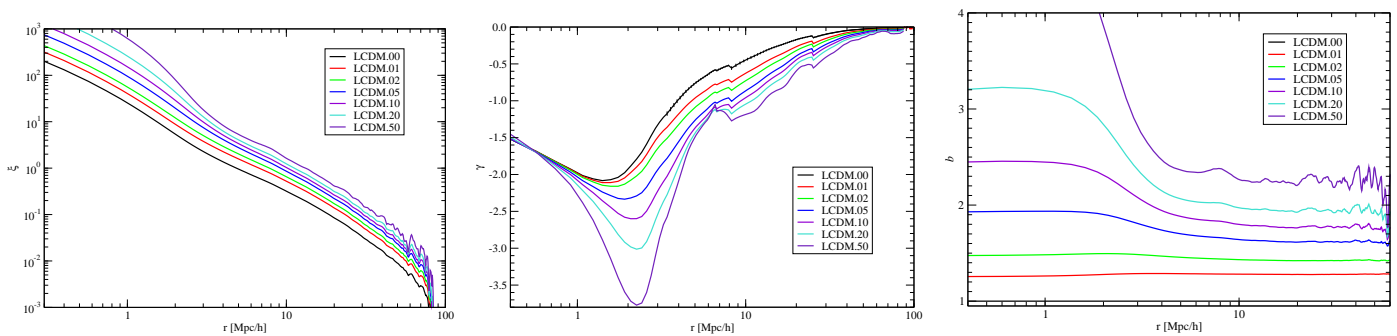
 
\centering 
\hspace{1mm}  
\resizebox{0.32\textwidth}{!}{\includegraphics*{LCDM-CF-rD.eps}}
\hspace{1mm}  
\resizebox{0.32\textwidth}{!}{\includegraphics*{LCDM-gamma-mod.eps}}
\hspace{1mm}  
\resizebox{0.32\textwidth}{!}{\includegraphics*{LCDM_3D_realCFbias.eps}}
\caption{Three-dimensional correlation and related functions of LCDM
  models for different particle density threshold $\delta_0$ limits.
  {\em Left panel:} CF. {\em Central panel}: Gradient function,
  $\gamma(r)= {d \log g(r) / d \log r}$.  {\em Right panel:} Bias
  function, $b(r, \rho_0) = \sqrt{\xi_C(r,\rho_0)/\xi_r(r)}$, using
  the CF of the full DM LCDM.00 model as a reference.  As argument we
  use the pair separation in 3D real space,
  $r=\sqrt{(\Delta\,x)^2+(\Delta\,y)^2+(\Delta\,z)^2}$. }
\label{fig:Fig1} 
\end{figure*}

{As shown by \citet{Einasto:2020aa}, and presented in the right panel
  of Fig.~\ref{fig:Fig1}, relative 3D CFs on medium scales, namely
  $4 \le r \le 20$~\Mpc, have a plateau similar to the plateau of relative
  power spectra around the wavenumber $k \approx 0.03$~$h$~Mpc$^{-1}$
  \citep{Einasto:2019aa}. We use the value of the CF at $r=6$~\Mpc\
to calculate the amplitudes of CFs, $A= \xi(6)$. This parameter is
well suited to characterising the amplitude of CFs, and is used here
instead of the correlation length, $r_0$.   We note that
  \citet{Einasto:1991oy} defined the amplitude of CFs using the value
  of CF at $r=1$~\Mpc.  The present study shows that, at this small
  separation, the presence of halos (clusters) distorts the smooth run
  of the CF, and therefore a separation of $r=6$~\Mpc\ is a better  value with which to
  characterise the CF amplitude.}

\subsection{Calculation of projected  CFs}

In the distant observer approximation, we can  use model data in
rectangular spatial coordinates.  In this approximation  we calculated  projected
CFs from 2D density fields using the \citet{Szapudi:2005aa} method.   First
we calculated 2D density fields on a  $2048^2$ grid by integrating the 3D
field, $\delta(x,y,z)$, in the $z$-direction (by ignoring
$z$-coordinates in the calculation of density fields in the selected
$z$ range) : 
\begin{equation}
  \delta_2(x,y) = \int_{z_1}^{z_2}  \delta(x,y,z) \dd{z}.
  \label{2dens}
\end{equation}
We integrated cubic samples to get $n$ sequentially
located 2D sheets of size $L_0 \times L_0 \times L$~\Mpc, where
$L = L_0/n$ is the thickness of the sheets, and $n = 1,~2,~4, \dots 2048$
is the number of sheets.  It is clear that $n=1$ corresponds to the
whole sample in the $z$-direction of thickness, $L=L_0=512$~\Mpc, $n=2$
corresponds to thickness $512/2 = 256$~\Mpc, and $n=2048$ corresponds
to thickness $L=512/2048=0.25$~\Mpc.  For each $n$ we calculated 2D
CFs for all $n$ sheets, and then found the mean CF and its error for
a given $n$.

Examples of 2D density fields for various thickness $L$ are given in
Fig.~\ref{fig:Fig9} below.  The upper panels of this figure show
sheets of thickness $L=8$~\Mpc\ at various $z$ locations and
demonstrate the variance of the cosmic density field.  The number of
particles/galaxies in sheets is large, and errors are small; for some
samples errors are shown in Fig.~\ref{fig:Fig6}.  In calculations, we
used the mean density of the whole field for a given particle density
limit $\rho_0$ (absolute magnitude $M_r$ limit). We also tested the
case in which we used the mean density of each sheet separately for
calibration.  This increases CF errors, but has little influence on
the amplitude of CFs.

Correlation functions were found using $L_{max}=200$~\Mpc\ for
90 logarithmic distance bins.  We label mean 2D sheets as LCDM.$i,n$
and Mill.$i,n$, where $i=\rho_0$ is the particle density limit used in
the selection of particles for LCDM samples (absolute magnitude $M_r$
limit for Millennium samples), and $n$ is the number of sheets in the
$z$-direction used to select particles (galaxies) for 2D samples. 
Millennium samples were analysed in real space and in redshift
space. We denote these samples as Millr.$i,n$ and Mills.$i,n$, 
respectively. Most of the analysis was made for Millennium samples in
real space. For simplicity, we refer to these samples as Mill.$i,n$
without the index $r$.

\section{Comparison of spatial and projected correlation functions}

In this section, we compare spatial CFs with respective projected
functions. We focus on studying the influence of the
thickness of 2D sheets on the behaviour of CFs. Also, we compare CFs in
real and redshift space.

\subsection{Three-dimensional  correlation  functions of LCDM models}

LCDM model samples are based on all particles of the simulation and
contain detailed information about the distribution of matter in regions
of different density.  As parameters, we use a particle density limit
$\rho_0$ in LCDM samples. For the argument in 3D CFs, we use the pair
separation in 3D space,
$r=\sqrt{(\Delta\,x)^2+(\Delta\,y)^2+(\Delta\,z)^2}$.  {Three-dimensional CFs,
  gradient, and bias functions of LCDM models are shown in Fig.~\ref{fig:Fig1}
  for a set of particle density limits $\rho_0$, as given in
  Table~\ref{Tab1}. }

Three-dimensional CFs behave as expected.  Samples with higher particle density
limit, which correspond to brighter galaxies, have CFs with higher
amplitudes.  This is a well-known effect, and was detected by
\citet{Bahcall:1983uq} and \citet{Klypin:1983fk}, and explained by
\citet{Kaiser:1984} as the biasing phenomenon.  The dependence of the
correlation amplitude and the respective correlation length on galaxy
luminosity has been the subject of many subsequent investigations. Among
these studies, we mention here the work by 
\citet{Norberg:2001aa}, \citet{Zehavi:2005aa,Zehavi:2011aa}, and
related studies of the power spectra of galaxies based on recent
large surveys such as \citet{Tegmark:2002aa}, \citet{Tegmark:2004aa}.

{The mean shape of CFs is traditionally characterised by the slope
  of the CF, which is given in the last column  of Table~\ref{Tab1}.  For small
  particle density limits of $\rho_0 \le 5,$ the slope is close to the
  traditional value, $\gamma \approx -1.8$, but for higher particle
  density limits the slope deepens.  Here we see the influence of DM
  halos.  A more detailed view of the shape of CFs is given   }
by the gradient function,
$\gamma(r)= {\dd{\log g(r)} / \dd{ \log r}}$, presented in
the central panel of Fig.~\ref{fig:Fig1}.  Here we clearly  see two regimes
of the CFs: on small scales, the CF describes the distribution of
particles (galaxies) in halos (clusters); on large scales it describes the distribution of
particles in the
cosmic web.  {The presence of two regimes in CFs and their
  interpretation as the transition from clusters to filaments was
  first noticed by \citet{Zeldovich:1982kl}, and was discussed in detail
  by \citet{Einasto:1992aa}.  The presence of a minimum in the
  gradient function and its interpretation as the effective radius of
  halos of the cosmic web was noted by \citet{Zehavi:2004aa}.
  The outer radius of halos can be identified with the minimum of the
  gradient function near $r \approx 2$~\Mpc.

  Dark matter halos of very different masses have almost identical DM
  profiles \citep{Wang:2020aa}.  As discussed by
  \citet{Einasto:2020aa}, at small separations the  gradient
  function measures the mean profile of all DM halos.  Near halo
  centres, the logarithmic gradient of DM halos is $-1.5$, and
  decreases to $-3.0$ at the periphery of halos
  \citep{Wang:2020aa}. This explains the constant value of the
  gradient, $\gamma(0.5)=-1.5$ at $r=0.5$~\Mpc, followed by a minimum
  near $r \approx 2$~\Mpc. For a small particle density limit of
  $\rho_0 < 5$, small-mass DM halos dominate, and the minimum is not
  deep.  With increasing particle density limit $\rho_0$, low-density
  halos are excluded from the sample and more massive halos
  dominate. This leads to an increase in the depth of the minimum of
  $\gamma(r)$. More massive halos have larger radii, thus the location
  of the minimum of the $\gamma(r)$ function shifts to higher
  separation $r$ values.  } It should be noted that a clear minimum is
present only in LCDM models.

{On larger scales, $r \ge 2$~\Mpc, the gradient function defines
  the fractal dimension of the sample, $D(r)=3+\gamma(r)$.  Our data
  show that the characteristic fractal dimension $D(r)$ is a function
  of separation $r$. On very large scales, the fractal dimension
  approaches the value $D(r)=3+\gamma(r) \rightarrow 3$,
  which is characteristic of a random distribution of particles/galaxies. }

\begin{figure*}[ht]
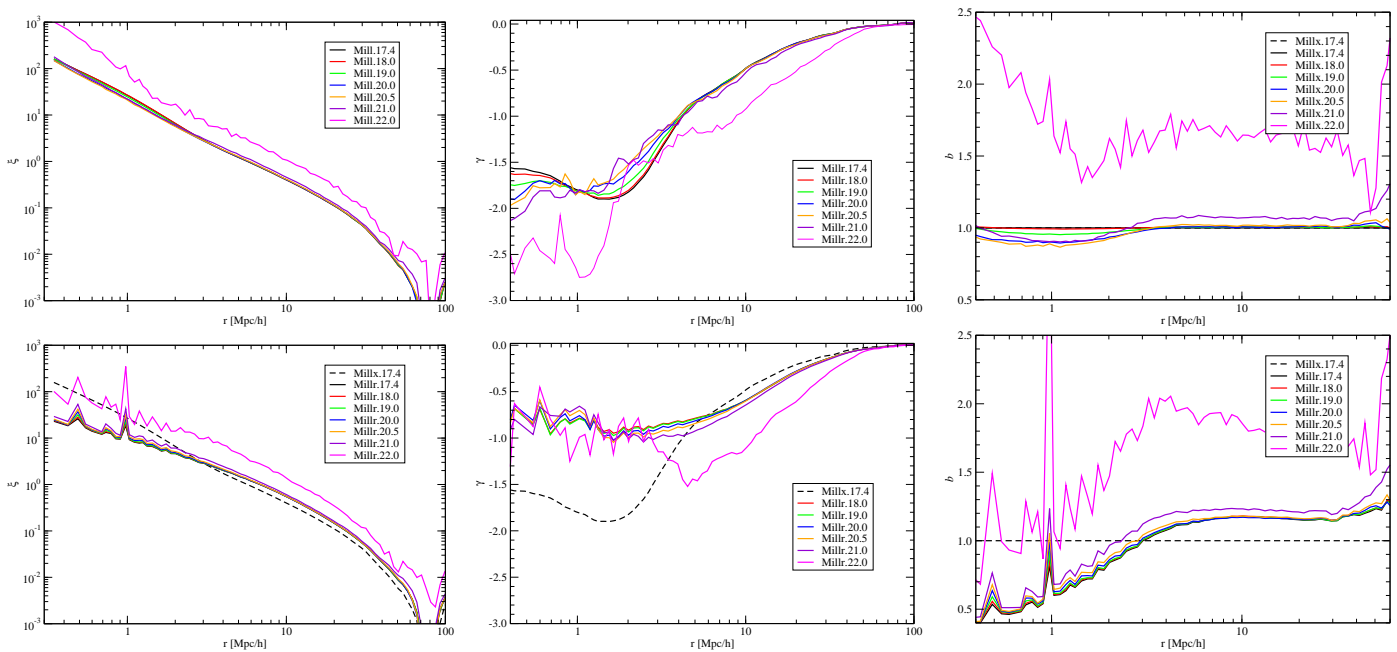
 
\centering 
\hspace{1mm}  
\resizebox{0.32\textwidth}{!}{\includegraphics*{Mill_3D_CF.eps}}
\hspace{1mm}  
\resizebox{0.32\textwidth}{!}{\includegraphics*{Mill_3D_CF_gamma.eps}}
\hspace{1mm}  
\resizebox{0.32\textwidth}{!}{\includegraphics*{Mill_3D_realCFbias.eps}}\\
\hspace{1mm}  
\resizebox{0.32\textwidth}{!}{\includegraphics*{Mill_3D_z_CF.eps}}
\hspace{1mm}  
\resizebox{0.32\textwidth}{!}{\includegraphics*{Mill_3D_CFred_gamma5.eps}}
\hspace{1mm}  
\resizebox{0.32\textwidth}{!}{\includegraphics*{Mill_3D_z_CFbias.eps}}
\caption{Three-dimensional correlation and related functions of Millennium models.
  {\em Top panels:}  Real space. {\em Bottom panels}:
  Redshift space.  {\em Left panels:}  CFs. {\em Central panels:}
  Gradient functions. {\em Right panels:} Relative bias
  functions.  For comparison we show  the 3D real space CF, the
  gradient, and the bias function of the sample 
  Mill.17.4  with dashed
  lines in the bottom panels.  As argument we use the pair separation in 3D real
  space. }
\label{fig:Fig2} 
\end{figure*}

{The right panel of Fig.~\ref{fig:Fig1} shows the bias function
  for the LCDM samples, with various particle density limits $\rho_0$.
  The limit $\rho_0=5$ brings the percolation properties of the sample
  LCDM.05 close to the percolation properties of the faintest galaxies of the
  SDSS survey, $M_r=-19.0$; the limit $\rho_0=10$ corresponds to
  $L^\ast$ galaxies of luminosity $M_r=-20.5$; see
  \citet{Einasto:2019aa}. For small particle density limits,
  $\rho_0 \le 2$, lines of bias function $b(r)$ are almost constant.
  In these samples, the smoothly distributed DM in low-density regions
  dominates the bias function.  In models with higher particle density
  limits, $\rho_0 \ge 5$, at small separations, $r \le 5$~\Mpc, the
  particle distribution in halos dominates the bias function. On
  larger separations, $r \ge 5$~\Mpc, bias functions are at an
  approximately constant level, and differ only by the amplitude $A$.}

\subsection{Three-dimensional correlation functions in real  space in
  Millennium models} 

{For
  Millennium samples, we used the \citet{Szapudi:2005aa} method to
  calculated  CFs and their derivates, gradient functions, 
  $\gamma(r)$, and relative bias function, $b_R(r)$.  Functions were
  calculated for absolute magnitude limits $M_r$, as given in
  Table~\ref{Tab2}.  In calculations with true model spatial positions
  $x,y,z$, we get functions in real space; see top panels of
  Fig.~\ref{fig:Fig2}. }

{The slopes of the CFs in the Millennium samples are close to the
  conventional value, $\gamma \approx -1.8$, for all magnitude $M_r$
  limits. LCDM samples behave differently and have higher negative
  slopes for samples with high $\rho_0$ values.  The reason for this
  difference is the weakness of clusters in Millennium samples.  This
  difference between LCDM and Millennium samples is more pronounced in
  gradient functions $\gamma(r)$.  In Millennium samples, only the
  sample Mill.17.4 has a clear minimum around 2~\Mpc; samples with
  higher luminosity limits have no minimum at this separation.  The
  sample Mill.17.4 contains galaxies that are faint enough to populate
  clusters of galaxies with fainter members.  Millennium samples with
  a higher luminosity limit contain less fainter cluster members,
  which makes the internal structure of the clusters less
  visible. {  In
  the samples with the highest luminosity limit, poor clusters
  disappear from the sample.  In the remaining clusters only their
  main galaxies are visible, and therefore  the correlation analysis considers
  these galaxies as isolated galaxies. }}

\begin{figure}[ht] 
\centering 
\hspace{1mm}  
\resizebox{0.40\textwidth}{!}{\includegraphics*{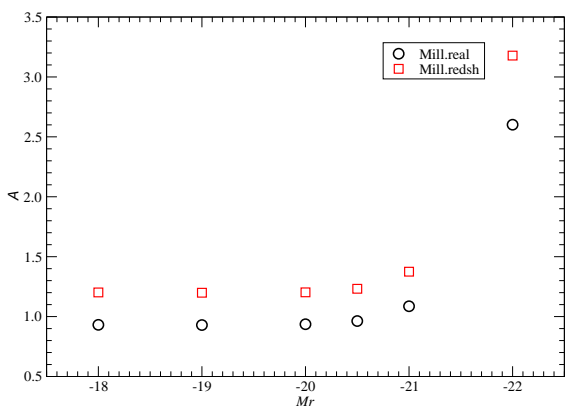}}
\caption{Amplitudes $A$ of the Millennium simulation CFs for various
  luminosity limits $M_r$. Real space and 
  redshift space CF amplitudes are shown using black
  and red symbols, respectively.  }
\label{fig:Fig3} 
\end{figure}

An important property of CFs is the dependence of  their amplitude on
particle density or luminosity limit $M_r$.  {The comparison of
  Figs.~\ref{fig:Fig1} and \ref{fig:Fig2} shows that the dependence of
  CF amplitudes of Millennium galaxy samples on the luminosity limit
  is different from the dependence of LCDM CF amplitudes on the
  particle density limits.  The amplitudes of the CFs of the LCDM samples increase
  approximately proportionally to the particle density limit of samples.
  In Millennium samples the amplitude $A$ remains almost constant with increasing
  $M_r$  up to $M_r=-20.0$.  This dependence is shown
  in Fig.~\ref{fig:Fig3}.  A flat distribution of correlation radii
  with increasing luminosity has been found in galaxies by
  \citet{Einasto:1991oy}, \citet{Norberg:2001aa},
  \citet{Zehavi:2005aa, Zehavi:2011aa}, and \citet{Einasto:2019aa}.
  This flat distribution of amplitudes of galaxy CFs can be explained
  by the absence of formation of very faint dwarf galaxies in voids.
  Dwarf galaxies form only as satellites in halos of brighter
  galaxies. This explains the almost constant amplitude of the CFs of
  galaxies of low luminosity, $M_r > -20.0$.}

{For Millennium galaxies, we calculated relative bias functions, 
  $b_R(r)=\sqrt{\xi(r)/\xi_0(r)}$, where $\xi(r)$ is the CF of the
  test sample, and $\xi_0(r)$ is the CF of the reference sample.  In
  the top right panel of Fig.~\ref{fig:Fig2} we show relative bias
  functions of the Millennium model using CFs of Millennium samples
  of galaxies for various luminosity limits $M_r$ in real space, and
  the CF of the sample Mill.17.4 in real space for reference.  For
  Millennium 
  samples, the CF of all DM is not available, and therefore it is possible to
  calculate only relative bias functions. Comparison of the right
  panels of Figs.~\ref{fig:Fig1} and \ref{fig:Fig2} shows the difference.}

\subsection{Three-dimensional CFs in  redshift space of Millennium models}

{We calculated CFs and gradient functions for Millennium samples
  in redshift space with galaxy positions in $x,y$ unaffected, but as
  $z$ positions we used redshift-distorted $z$ coordinates, as seen by
  a distant observer. Redshift space CFs and related functions are
  shown in the bottom panels in Fig.~\ref{fig:Fig2}.}

The almost constant level of amplitude of CFs is valid both in real
and redshift space; see Table~\ref{Tab2} and  Figs.~\ref{fig:Fig2} and \ref{fig:Fig3}.  This
property can be quantified by the  amplitude correction factor of redshift
CFs,
\be
C= \sqrt{A_s/A_r},
\label{bias1}
\ee
where $A_s=\xi(6)_s$ is the CF amplitude of a sample at 6~\Mpc\ in
redshift space, and $A_r=\xi(6)_r$ is the CF amplitude at 6~\Mpc\ of the
sample in real space.  Amplitude correction factors $C$ of Millennium
galaxy samples in redshift space are given in   Table
\ref{Tab2}.  As we see, for fainter samples $M_r > -20.0$ the 
amplitude correction factor  $C$ is almost constant with the mean value
$C = 1.135 \pm 0.001$. This value is expected from the \citet{Kaiser:1987aa}  effect ---
the contraction of superclusters in the vertical direction is, $C^2=g$,
where $g$ is calculated from Eq.~(\ref{Kaiser}) below.

\begin{figure}[ht] 
\centering 
\hspace{1mm}  
\resizebox{0.40\textwidth}{!}{\includegraphics*{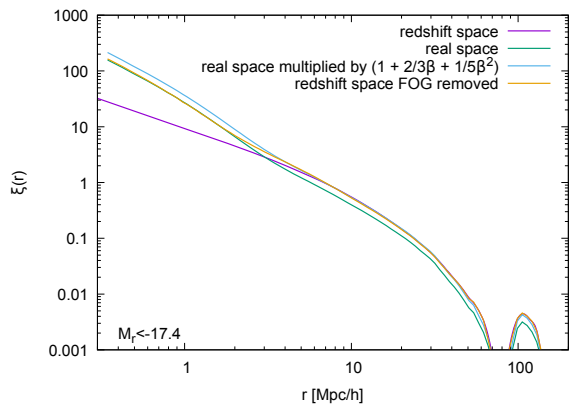}}
\caption{Three-dimensional CFs of Mill.17.4  samples in real space, redshift space,
  Kaiser space, and redshift space with the FOG removed. 
  }
\label{fig:Fig4} 
\end{figure}

\begin{figure}[ht] 
\centering 
\hspace{1mm}  
\resizebox{0.2\textwidth}{!}{\includegraphics*{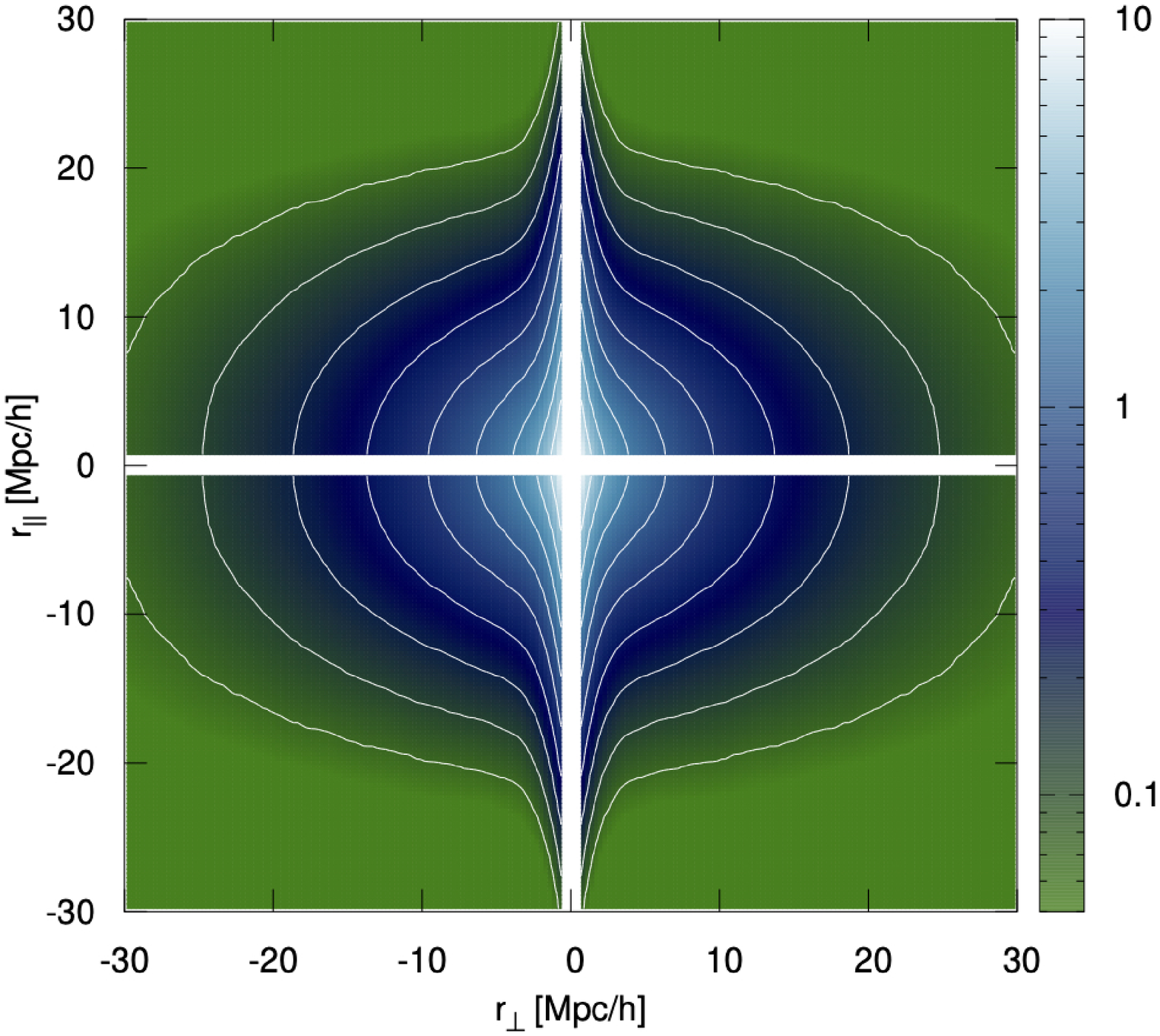}}
\hspace{1mm}  
\resizebox{0.2\textwidth}{!}{\includegraphics*{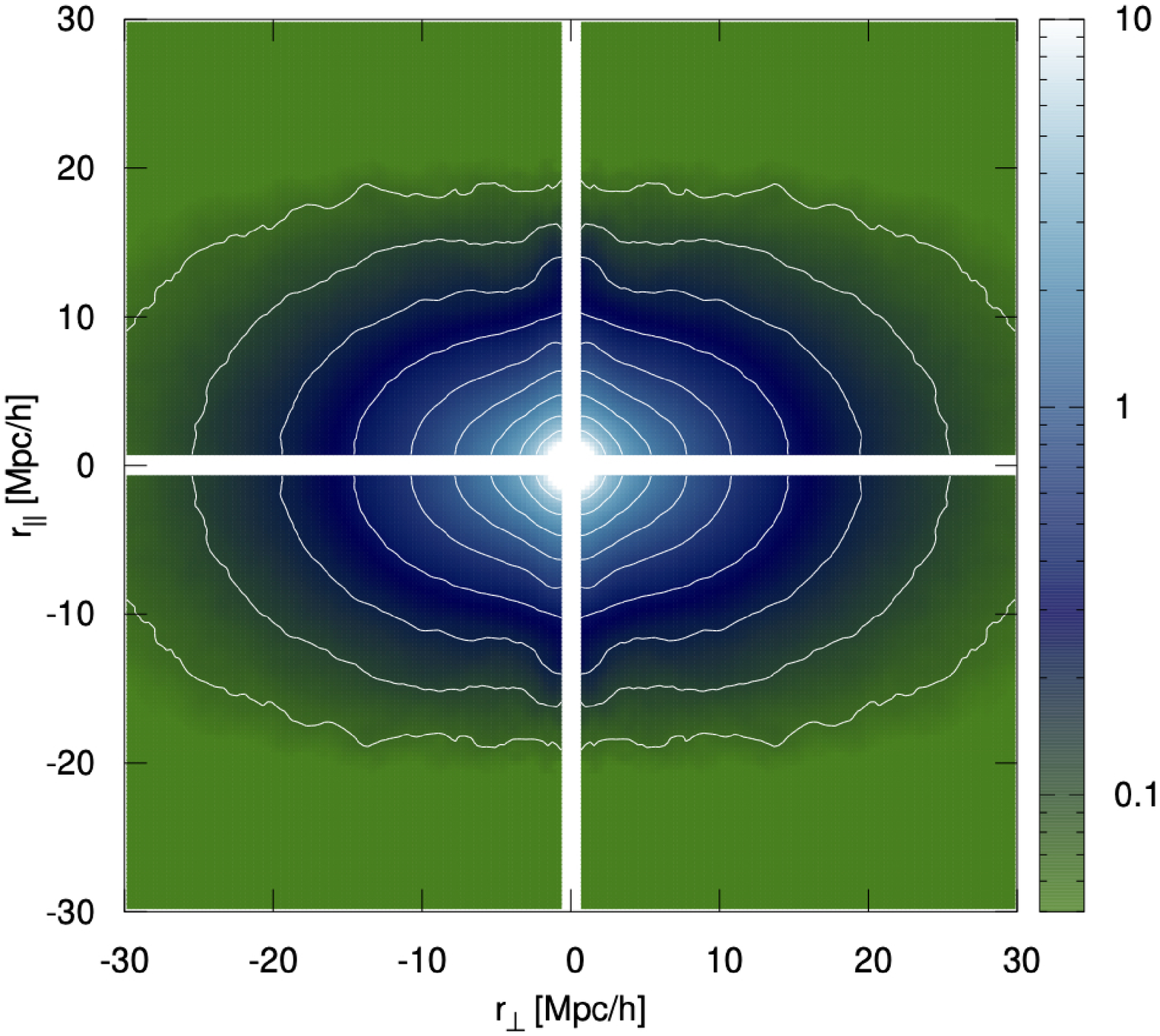}}
\caption{Two-dimensional  anisotropic  CFs of Mill.17.4  samples in  redshift space
  (left) and in redshift space with the FOG  removed using the procedure
  described in the main text (right). 
  }
\label{fig:Fig5} 
\end{figure}

The comparison of CFs shows that the amplitudes of
Millennium samples in redshift space at small separations are much
lower.  This has a simple explanation: In redshift space, clusters of
galaxies are expanded in the $z$-direction by velocity dispersions, and
lose their compact character.  This decreases the number of close
pairs at small separations, and in this way decreases the amplitude of
CFs. For this reason, CFs at small separations are shallower, and have a
lower negative gradient $\gamma(r)$; see Fig.~\ref{fig:Fig2}. Galaxies
from different clusters are partly mixed; this increases the scatter
of the CF and its gradient function.

At large separations, the  amplitudes of CFs in redshift space are
 larger. Here, the \citet{Kaiser:1987aa} effect is visible: in
redshift space massive supercluster-type systems are contracted in the
$z$-direction; their effective volume decreases, and the volume of
voids between superclusters increases. This leads to the increase of
the amplitude of CFs.

\begin{figure*}[ht]
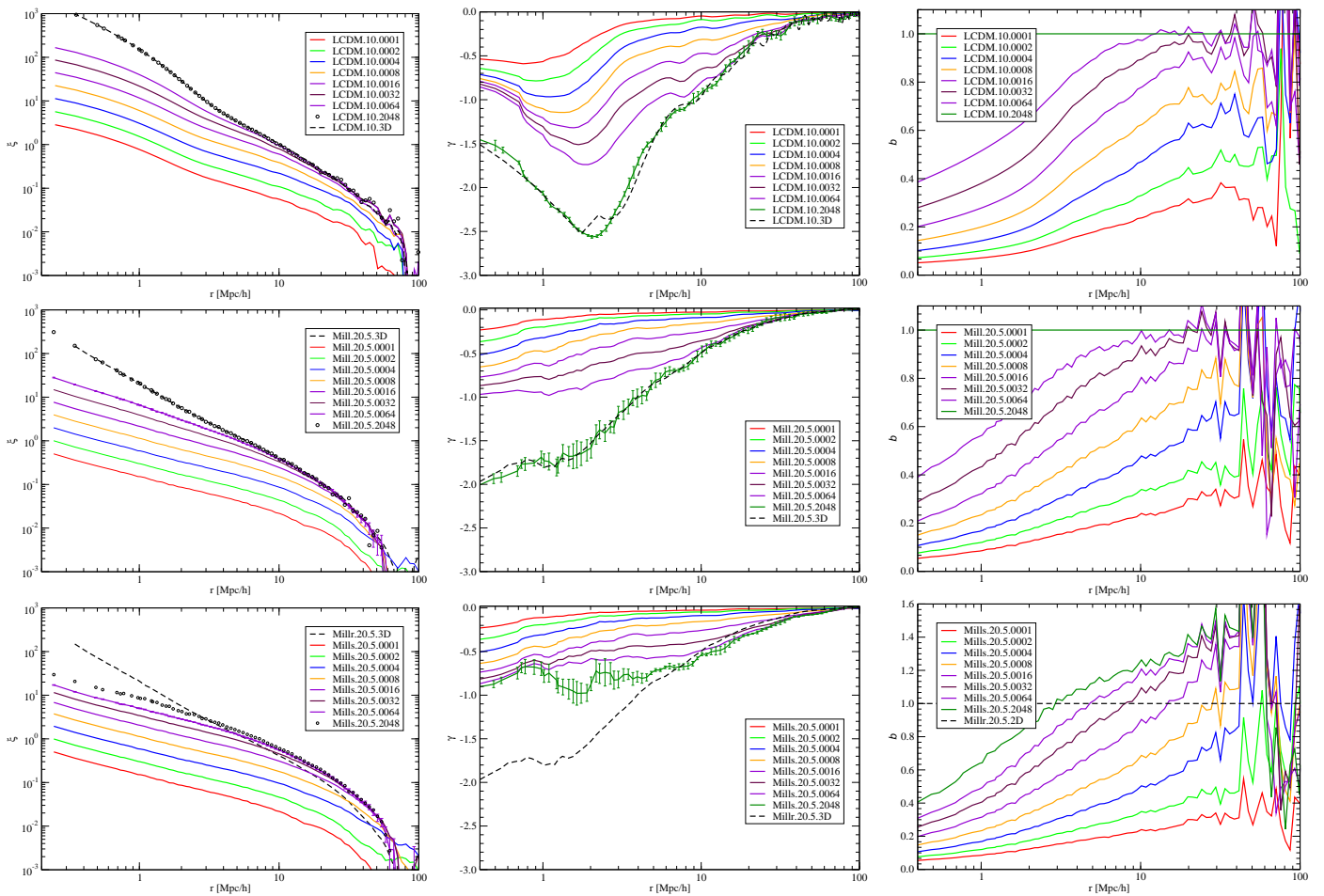
 
  \centering 
\hspace{1mm}  
\resizebox{0.32\textwidth}{!}{\includegraphics*{cf_2D_10.eps}}
\hspace{1mm}  
\resizebox{0.32\textwidth}{!}{\includegraphics*{cf_2D_10_gamma.eps}}
\hspace{1mm}  
\resizebox{0.32\textwidth}{!}{\includegraphics*{cf_2D_10_bias.eps}}\\
\hspace{1mm}  
\resizebox{0.32\textwidth}{!}{\includegraphics*{Mill_3D_2D_20.5.eps}}
\hspace{1mm}  
\resizebox{0.32\textwidth}{!}{\includegraphics*{Mill_2D_10_gamma.eps}}
\hspace{1mm}  
\resizebox{0.32\textwidth}{!}{\includegraphics*{Mill_2D_10_bias.eps}}\\
\hspace{1mm}  
\resizebox{0.32\textwidth}{!}{\includegraphics*{Mill_3D_z-2D_20.5.eps}}
\hspace{1mm}  
\resizebox{0.32\textwidth}{!}{\includegraphics*{Mill_2D_10_z_gamma.eps}}
\hspace{1mm}  
\resizebox{0.32\textwidth}{!}{\includegraphics*{Mill_2D_10_z_bias.eps}}
\caption{Two-dimensional CFs and related functions for models for
  different thickness of 2D samples. { {\em Top row:} LCDM model with
  particle density limit $\rho_0=10$. {\em Second row:} Millennium
  samples with magnitude limit $M_r=-20.5$ in real space. {\em Bottom
    row:} Millennium samples with magnitude limit $M_r=-20.5$ in
  redshift space.  {\em Left panels}: CFs. {\em Central panels}:
  Gradient functions. {\em Right panels}: Relative bias functions. } As
  argument we use pair separations perpendicular to the line of sight,
  $r_p= \sqrt{(\Delta\,x)^2+(\Delta\,y)^2}$. As parameter we use the
  thickness of samples, $L$.  Lines of various colours mark 2D samples
  of different thickness.  For comparison we show 3D functions for
  samples with $\rho_0=10$ and $M_r=-20.5$. Error bars are shown for
  some samples. }
\label{fig:Fig6} 
\end{figure*} 

We calculated CFs in redshift space with FOG removed by compressing
FOG to get equal dispersions of cluster galaxies in radial and
tangential directions; a similar procedure was applied by
\citet{Einasto:1986ab} and \citet{Tegmark:2004aa}.  The compression
rate was found empirically without any assumptions regarding the underlying
cosmology.  Calculated CFs of the sample Mill.17.4 are shown in
Fig.~\ref{fig:Fig4}.  The expected CF in redshift space can be found
by multiplying the real space CF by a factor  \citep{Kaiser:1987aa}:
\be
g=1+ 2/3\,\beta +1/5\,\beta^2,
\label{Kaiser}
\ee
where $\beta= \Omega_m^{0.55}/b$, and $b$ is the bias factor.   We used
$\Omega_m=0.25$ (value used in Millennium simulation) and 
 bias factor $b=1$.  Figure ~\ref{fig:Fig4} shows that at $r \ge 5$~\Mpc,\ the CF in
redshift space, expected from real space and using this conversion
factor, coincides with the actual redshift space CF.

{For comparison, we also calculated anisotropic two-point CFs for
 Millennium galaxies with magnitude limit $M_r=-17.4$ in redshift
 space,  and in redshift space with FOG removed using the procedure
 described above; see Fig.~\ref{fig:Fig5}. The right panel presents anisotropic 2D CF of
 Mill.17.4 galaxies in Kaiser space. As we see, FOGs are essentially
 removed; however  some remnants are still visible.  In spite of these
 remnants, the 3D CF of Mill.17.4 sample in redshift space with FOG
 removed coincides at small distances $r \le 2$~\Mpc\ with the 3D CF in
real space.  At larger distances, it coincides with the 3D redshift CF; see
Fig.~\ref{fig:Fig4}. 

The bottom right panel of Fig.~\ref{fig:Fig2} shows relative bias
functions for Millennium galaxies in redshift space. Here we also use
the sample Mill.17.4 in real space as reference. This function is
determined by redshift distortions: On small scales, $r \le 3$~\Mpc,
the FOG effect dominates; on larger scales the contraction of
superclusters by the Kaiser effect dominates.  }

{The bottom left and right panels of Fig.~\ref{fig:Fig2} show that CFs
  and relative bias functions have peaks at $r=1$~\Mpc\ and
  elsewhere. To find the meaning of the peaks we made additional
  calculations of redshift space CFs using various resolutions in the
  \citet{Szapudi:2005aa} method from $N_{\mathrm{grid}} = 128$ to
  $N_{\mathrm{grid}} = 3072$.  These calculations showed that the
  effect is due to anisotropy and interference of redshift corrections
  on various scales, and each grid size generates peaks at different
  scales. }

\subsection{Two-dimensional correlation functions}

{We calculated 2D CFs, gradient functions, and relative bias functions
  for the LCDM and Millennium samples using the \citet{Szapudi:2005aa}
  method, as described above.  Results are shown in
  Fig.~\ref{fig:Fig6} for a series of sample thicknesses
  $L=L_0/n$~\Mpc, using number of sheets
  $n=1,~2,~4,~8,~16,~32,~64, \text{and }~2048$, which correspond to
  sheet thicknesses from the maximum, $L=512$~\Mpc\ (number of sheets
  $n=1$), to the mean of $n=2048$ most thin sheets, each of a
  thickness of $L=0.25$~\Mpc.  { We use LCDM.10 samples with particle
  density limit $\rho_0=10$, and Millennium samples Mill.20.5 with
  luminosity limit $M_r=-20.5$.}  Both limits correspond approximately
  to $L^\ast$ galaxies \citep{Einasto:2019aa}.  As argument in 2D CFs
  we use the pair separation perpendicular to the line of sight for a
  distant observer, $r_p= \sqrt{(\Delta\,x)^2+(\Delta\,y)^2}$.}

Two-dimensional CFs depend on two parameters, the thickness of the sheets, $L = 512/n$~\Mpc,
and the particle density limit of LCDM samples, $\rho_0$, or magnitude
limit, $M_r$, of Millennium samples.  {The essential properties of the CFs
  are their amplitudes, $A$, and shape
  characteristics: slope $\gamma$, and  gradient and relative bias
  functions.  
Correlation amplitudes of 2D samples of LCDM models are given in Table
\ref{Tab3}, and correlation amplitudes of 2D samples of Millennium
models in real space are given in Table \ref{Tab4}.  Correlation
amplitudes for full 2D samples LCDM.00 are printed in italics in the tables.  For
comparison, we provide the  amplitudes of 3D CFs for
various particle density limits  in the last column, the amplitude of the 3D
CF of the model LCDM.00 in boldface, that is, ~the full DM LCDM model.  We calculated
amplitudes of 2D Millennium samples also in redshift space; these are
used in the bottom panels of Fig.~\ref{fig:Fig6}.}

\subsubsection{Two-dimensional correlation functions of LCDM model samples }

\begin{figure}[ht] 
\centering 
\hspace{1mm}  
\resizebox{0.40\textwidth}{!}{\includegraphics*{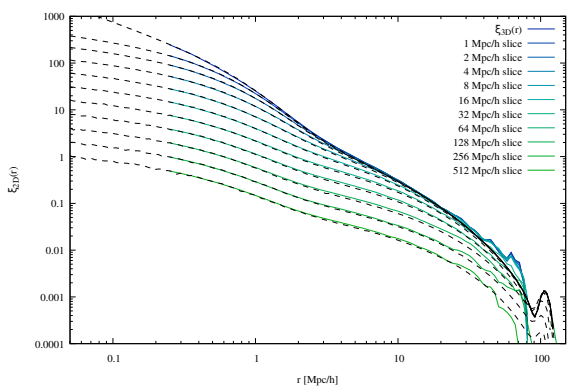}}
\caption{Two-dimensional CFs of LCDM models, calculated from 2D density fields
  (solid coloured lines), and analytically from 
  the 3D CF model using Eq.~(\ref{eq05}) (dashed lines).
  }
\label{fig:Fig7} 
\end{figure}

\begin{figure*}[ht]
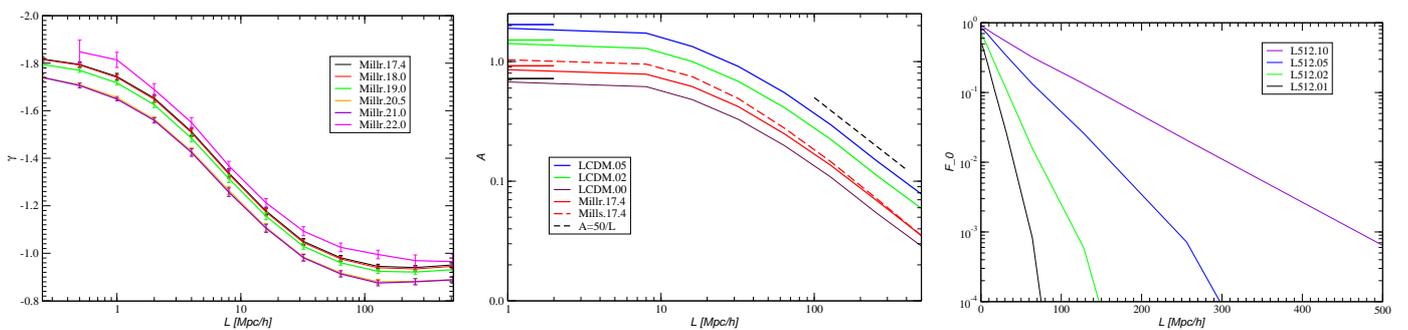
 
\centering 
\hspace{1mm}  
\resizebox{0.32\textwidth}{!}{\includegraphics*{Mill_2D_slope-L.eps}}
\hspace{1mm}  
\resizebox{0.32\textwidth}{!}{\includegraphics*{LCDM_Mill_Ampl-L.eps}}
\hspace{1mm}  
\resizebox{0.32\textwidth}{!}{\includegraphics*{L512_2D_F0-Lb.eps}}
\caption{ {\em Left  panel:}  Dependence of the slope
  $\gamma$ of Millennium 2D CFs {of various magnitude limits} on the thickness of samples $L$.
  {\em Middle panel}: Amplitudes $A=\xi(6)$ of several LCDM and
  Millennium simulation 2D CFs as functions of the thickness of
  samples, $L$. For the Mill.17.4 sample, amplitudes are shown for real
  space and redshift space CFs. Short lines show 3D CF amplitudes of
  the same samples. The black dashed line shows the approximation
  $A=50/L$.  
  {\em Right panel}: Fraction of  zero-density
  cells ($\rho \le 0.001$) in 2D density fields, $F_0$, as a function of
  the thickness of 2D samples, $L$.  }
\label{fig:Fig8} 
\end{figure*}

The luminosity dependence of the 2D CFs of  the LCDM model is similar to the luminosity
dependence of 3D CFs. With increasing luminosity, the amplitude of CFs
increases, as shown in Tables \ref{Tab1} and \ref{Tab2} for 3D
functions, and in Tables \ref{Tab3} and \ref{Tab4} and Fig.~\ref{fig:Fig8} for 2D functions.
{This means that 2D density fields and respective CFs preserve the
  information on the luminosity dependence contained in 3D density
  fields and 3D CFs. To check this,
  we calculated the 2D CFs of the LCDM.00 model analytically,
  applying Eq.~(\ref{eq05}), and using the 3D CF of a LCDM halo model.
  The results of this calculation are shown in Fig.~\ref{fig:Fig7}. As we
  see, the analytic model describes the decrease in amplitude of the 2D
  CFs with increasing sample thickness as well as the numerical
  integration by the \citet{Szapudi:2005aa} method.}

{\scriptsize 
\begin{table*}[ht] 
\caption{Amplitudes of 2D CFs of LCDM models.} 
\label{Tab3}                         
\centering
\begin{tabular}{lrccccccccc}
\hline  \hline
Sample   & $\rho_0$& 1&2 &4 & 8&16& 32&64&2048&3D \\  
& & 512&256&128&64&32&16&8&0.25&\\ 
\hline  
LCDM.00   &  0& {\em 0.0280}  &{\em 0.0540} &{\em 0.1084} &{\em 0.198}
                                 &{\em 0.329} &{\em 0.481} &{\em 0.617}
                                           &{\em 0.717}& {\bf  0.722 }\\    
LCDM.01   &  1& 0.0461   & 0.0889 &0.1784 &0.326 &0.541 &0.792 &1.017 &1.185&1.191\\
LCDM.02   &  2&0.0579   &0.1117 &0.224 &0.412 &0.683 &1.000 &1.289   &1.505&1.515\\
LCDM.05   &  5& 0.0764  &  0.1477 &0.297 &0.549 &0.910 &1.336 &1.730 &2.017&2.039\\
LCDM.10 &  10&0.0917 & 0.1781& 0.358 &0.667 &1.105 &1.620 &2.105 &2.430&2.478\\
LCDM.20 &  20& 0.1092 & 0.212 &0.430 &0.810 &1.337 &1.961 &2.558 &2.933&2.986\\
LCDM.50 &  50&0.1356 & 0.275& 0.574 &1.106 &1.803 &2.644 &3.467 &3.811&3.954\\
\hline 
\end{tabular} 
\tablefoot{Table columns starting from the left: sample name,
  particle-density limit $\rho_0$, 2D CF amplitudes, $A= \xi(6)$ for
  sample parameter $n=1,~2,~4,~8,~16,~32,~64,~2048$.  { The last
    column gives amplitudes from the 3D analysis.  In the table head,
    the upper line gives $n$, and the next line gives sheet thickness
    $L = 512/n$ in \Mpc. } }
\end{table*} 
}

{\scriptsize 
\begin{table*}[ht] 
\caption{Amplitudes of 2D CFs of Millennium models.} 
\label{Tab4}                         
\centering
\begin{tabular}{lrcccccclll}
\hline  \hline
Sample   & $M_r$& 1&2 &4 & 8&16& 32&64&2048&3D \\  
& & 512&256&128&64&32&16&8&0.25&\\ 
\hline  
Millr.17.4&-17.4 &0.0342 &0.0687 &0.1359& 0.248&  0.420&   0.618&   0.785&  0.906&   0.922\\
Millr.18.0&-18.0 &0.0347 &0.0697 &0.1374& 0.250&  0.424&   0.625&   0.791&  0.916&   0.931\\
Millr.19.0&-19.0 &0.0346 &0.0695 &0.1374& 0.250&  0.423&   0.622&   0.792&   0.911&  0.929\\
Millr.20.0&-20.0 &0.0349 &0.0696 &0.1381& 0.251&  0.424&   0.625&   0.796&   0.925&  0.936\\
Millr.20.5&-20.5 &0.0366 &0.0724 &0.1423& 0.257&  0.434&   0.638&   0.811&   0.966&  0.963\\
Millr.21.0&-21.0 &0.0408 &0.0812 &0.1597& 0.289&  0.483&   0.720&   0.925&   1.106&  1.086\\
Millr.22.0&-22.0 &0.1006 &0.1726 &0.3600& 0.692&  1.175&   1.662& 2.135& 2.2:  &  2.601\\
\hline 
\end{tabular} 
\tablefoot{Table columns starting from the left: sample name,
  particle-density limit $\rho_0$, 2D CF
  amplitudes, $A= \xi(6)$ for  sample parameter
  $n=1,~2,~4,~8,~16,~32,~64,~2048$. { The last
  column gives  amplitudes from 3D analysis. In the table head the
  upper line gives
  $n$, and the next line gives sheet thickness $L = 512/n$ in \Mpc. } }
\end{table*} 
}

{The gradient function, $\gamma(r)$, characterises the fine
  structure of the cosmic web. The top middle panel of
  Fig.~\ref{fig:Fig6} shows that the internal structure of the DM halos of
  the LCDM model is partially preserved in 2D CF gradient functions.
  As expected, this information is fully preserved in very thin 2D
  slices. The characteristic minimum of gradient functions near
  $r \approx 2$~\Mpc\ is visible, but has a lower amplitude.  The mean
  gradient of LCDM 2D CFs at small separations,
  $\gamma(r < 1) \approx -0.7$, and varies in large limits,
  $-1.7 \le \gamma(r) \le -0.5$ for separations $2 \le r \le 10$~\Mpc.
  The fine structure of DM halos is gradually erased with increasing
  thickness of 2D slices. At large separations, gradient 
  functions approach zero values, as expected for random samples.}

\subsubsection{Two-dimensional correlation functions of Millennium model samples}

{The general shape of CFs is described by the slope $\gamma$ of the
  CFs.  The comparison of Figs.~\ref{fig:Fig1}, \ref{fig:Fig2} and
  \ref{fig:Fig6} shows that 2D CFs are much shallower than 3D CFs.
  The slope of 2D CFs depends on the thickness of samples, $L$. This
  dependence is shown in the left panel of Fig.~\ref{fig:Fig8} for
  Millennium 2D samples in real space for various luminosity limits.
  For thick samples with $L \ge 32$~\Mpc,\ the slope is
  $\gamma \approx -0.9$, close to the characteristic slope of 2D CFs, which is well
  known from early studies of CFs: \citet{Peebles:1973a},
  \citet{Peebles:1975}, and \citet{Peebles:2001aa}.  With decreasing
  thickness of 2D samples, the negative slope increases, and for
very thin sheets  reaches a value of $\gamma \approx -1.8$, which is characteristic
  of 3D CFs. There is only a weak dependence of the slope on the  
  magnitude limit, $M_r$.  }

{Figure~\ref{fig:Fig6} shows that the information on the internal
  structure of the clusters of the Millennium samples is fully lost in 2D CFs
  of simulated galaxies of the sample Mill.20.5 (and of samples with
  brighter luminosity limit). At this luminosity threshold, clusters
  contain only a few galaxies.  In the 2D field, clusters are
  superposed by galaxies from different vertical locations in
  projection; see Fig.~\ref{fig:Fig9}.  The amplitudes of the 2D CFs of
  the Millennium samples are rather low (see Table 4), and in the calculation
  of the gradient function, the first constant term of the function
  $g(r)= 1 + \xi(r)$ dominates.  At small separations, the gradient
  function of the 2D Millennium CFs lies in the range
  $ -1.0 \le \gamma(r) \le -0.2$, depending on the thickness $L$.  For
  thick samples over the whole separation interval the gradient
  functions have values of $\gamma(r) > -0.5$, that is,\, the respective 2D CFs
  are similar to the CFs of samples with a random distribution of galaxies.}

{In the present context it is
  interesting that the difference between gradients of 2D and 3D CFs
  on small scales gradually fade away with decreasing thickness $L$
  of subsamples.  The mean 2D CF of very thin slices is almost equal
  to the 3D CF of the same sample, as seen in Fig.~\ref{fig:Fig6} and
  in Tables \ref{Tab3} and \ref{Tab4}.  This is expected, because thin  
  slices contain all essential spatial information of the 3D density
  field: the fraction of zero-density cells and the mutual distances of
  high-density regions. The last point is essential, as the CF is
  sensitive not to the location of galaxies/particles, but to their
  separations. Because of this property, the mean 2D CF of thin slices is
  equivalent to the 3D CF, in spite of the loss of information on the
  $z-$coordinates of galaxies/particles.  }

\subsubsection{Amplitudes of 2D correlation functions }

The middle  panel of Fig.~\ref{fig:Fig8} shows the amplitudes of the
2D CFs from three LCDM simulations 
and one Millennium simulation  as functions of the thickness of the
samples, $L$. For the Mill.17.4 sample, amplitudes are shown for real
and redshift space.  Short horizontal lines on the left axis show
amplitudes of 3D CFs of the same samples. {At small thickness
  values, lines for 2D model samples smoothly approach the limits given by
  3D samples.  } The figure shows that the decrease in the amplitude
with increasing thickness is very regular, and almost identical in all
samples relative to the amplitude of the 3D CF.  A decrease in the
amplitude of the 2D CF  by about 10\,\% is seen also at sheets of  thickness
$L=8$~\Mpc. At $L=512$~\Mpc,\ the amplitudes of 2D CFs are lower than
the amplitudes of 3D CFs by more than an order of magnitude.  Also, we see
that the difference in amplitudes of Millennium samples in real and
redshift space is very small.

{The amplitudes shown in Fig. ~\ref{fig:Fig8} were calculated
  numerically from 3D CFs using the \citet{Szapudi:2005aa} method as
  described above.  Integration using the analytical expression
  Eq.~(\ref{eq05}) yields identical results. As seen from
  Eq.~(\ref{eq07}),  according to the Limber approximation, 
  amplitudes are expected to be inversely proportional to thickness $L$.
  Figure~\ref{fig:Fig8} shows that the Limber approximation yields a
  good representation of the $A(L)$ function for large thicknesses,
  $L\ge 30$~\Mpc.}

 The most essential aspect of Figs.~\ref{fig:Fig6} and \ref{fig:Fig8} is their suggestion
 that the behaviour of the 2D CFs of  samples of different thickness
 of LCDM and Millennium model samples 
 is very similar: the amplitudes of 2D CFs depend significantly  on
 sample thickness and are {\em always} lower than the amplitudes of the 
3D CFs of the same sample.   The greater the thickness of the 2D
sample, the greater the difference between  the amplitudes of 2D and
3D CFs. This behaviour is almost identical in 
all our samples, in LCDM and Millennium samples in real space, and in
Millennium samples in redshift space.  This similarity is remarkable,
because test particles in samples are defined differently: DM particles
with local density labels and simulated galaxies with luminosity
labels. {This similarity means that the geometrical properties of the
  cosmic web are very similar in both models, defined in our LCDM
  model by the particle density limits, $\rho_0$, and in Millennium
  samples by the simulated galaxy luminosity limit, $M_r$.  The similarity
  in the geometrical properties of the density fields, defined by LCDM
  models and SDSS galaxies, was demonstrated by
  \citet{Einasto:2019aa} using the extended percolation analysis.
  Correlation and percolation analyses are sensitive to different
  aspects of the structure of the cosmic web, but yield similar results for the
  amplitude dependence on sheet thickness.  }

{The second essential aspect of Fig.~\ref{fig:Fig8} is the shape
  of the $A(L)$ function. At small thickness, $L \le 10$~\Mpc,
  the amplitudes of 2D CFs are approximately equal to the amplitudes of 3D
  CFs.  This means that the spatial structure of thin 2D slices is
  statistically similar to the spatial structure of the whole 3D
  web. Most importantly, the fraction of zero-density cells of the
  cosmic web is similar in thin 2D slices and in the whole 3D web.
  The fraction of zero-density cells, $F_0$ (actually cells of density
  $\rho \le 0.001$), is shown in the right panel of Fig.~\ref{fig:Fig8},
  also as a function of  2D sheet thickness, $L$.  We show the
  fraction $F_0$ for four particle density limits,
  $\rho_0 =1,~2,~5,~10$.  We see that, with increasing thickness, the
  fraction of zero-density cells in 2D slices decreases. The middle
  panel shows that, for thicknesses $L \ge 30$~\Mpc,\ the amplitude is
  inversely proportional to the thickness, $A \propto 1/L$.  In this
  thickness range, the decrease in the fraction of zero-density cells
  $F_0$ with increasing thickness $L$ of 2D slices is the deciding
  factor that determines the amplitude of 2D CFs.  We discuss this
  aspect  further in the following section.  }

\subsubsection{Relative bias functions of 2D CFs}

{The right panels of Fig.~\ref{fig:Fig6} show relative bias functions
  calculated for particle density limit $\rho_0=10$ for the LCDM model,
  and magnitude limit $M_r=-20.5$ for Millennium galaxy samples.  In
  calculations of relative bias functions we used the samples LCDM.10
  and Mill.20.5 (in real space)  as reference samples for the  
  3D CFs.  We show
  the dependence of relative bias functions on   sample thickness
 $L= 512/n$~\Mpc, where $n=1,~2,~4,~8,~16,~32,~62,~2048$ is
  the number of sheets in 2D density fields.

  Relative bias functions show the dependence of relative  CFs on
  the separation $r$.  Amplitudes $A$  were calculated for a fixed
  separation value, $r=6$~\Mpc.  The figure shows that the decrease in
  the bias function and amplitude with increasing sample thickness
  $L$ is regular, and  similar in LCDM and Millennium samples
  in real space.  All relative bias functions have values of less than
  unity, that is, 2D CFs have lower amplitudes than 3D CFs.  One  exception
  is the relative bias function of the Millennium samples in redshift
  space, shown in the bottom right panel of Fig 3.3. For thinner samples, $n
  \ge 8$, we see relative bias values of $b_R \ge 1$, which is caused by redshift
  distortions due to the Kaiser effect. This excess is seen also in
  CFs shown in the bottom left  panel.
}

\begin{figure*}[ht] 
\centering 
\hspace{1mm}  
\resizebox{0.95\textwidth}{!}{\includegraphics*{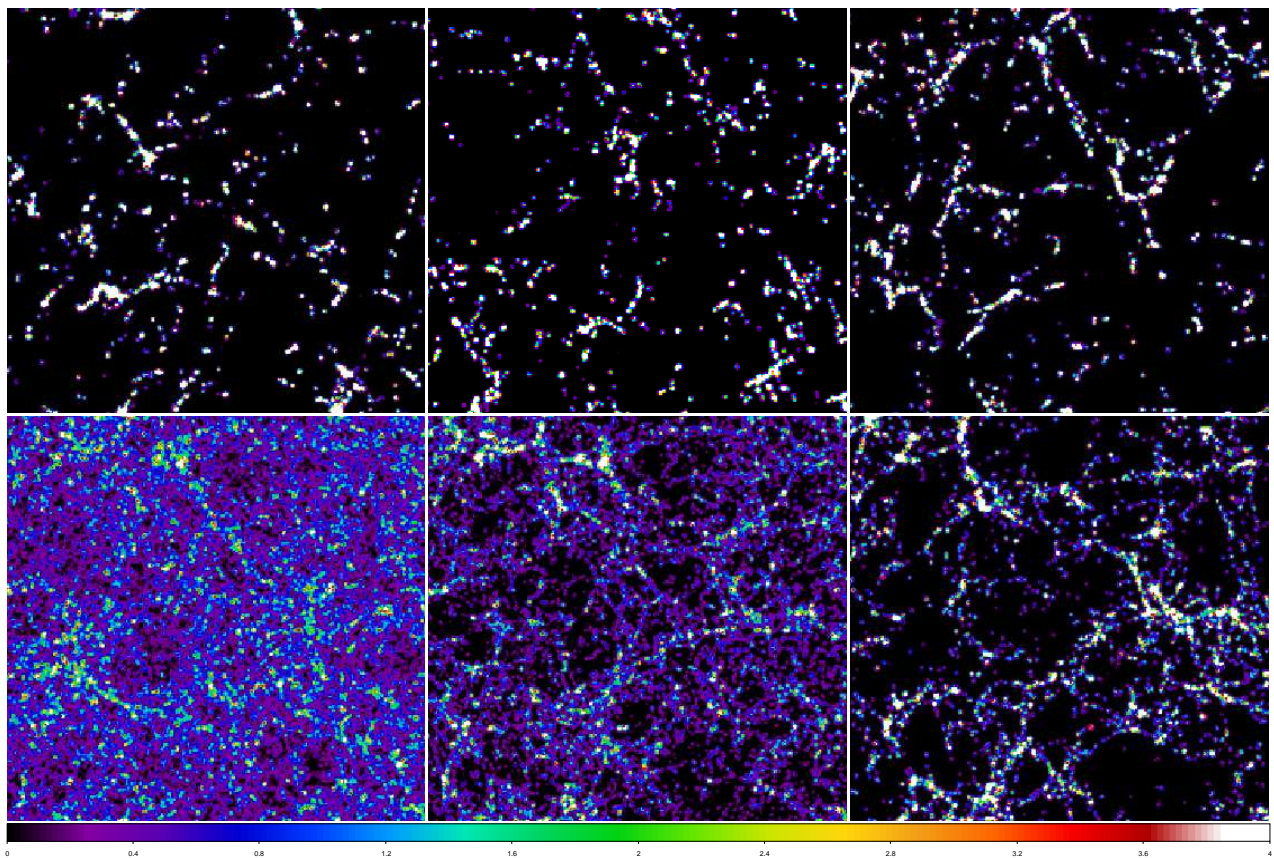}}
\caption{Two-dimensional density fields of LCDM.10 models with
  particle density threshold $\rho_0=10$. {\em Top panels:} 2D density
  fields of thickness $L=8$~\Mpc\ at {$z$-coordinates
    $100,~200,~300$.}  {\em Bottom panels} (from left to right): Thickness
  $L=512,~128,~32$~\Mpc.  We show the central sections of 2D density
  fields of size $256 \times 256$~\Mpc.  The colour scale is linear,
  and the code is identical in all panels.  }
\label{fig:Fig9} 
\end{figure*}

\section{Correlation functions as descriptors of the cosmic web}

Here, we describe spatial and projected density fields first, and their
description by CFs. Thereafter, we analyse CFs as descriptors of the
cosmic web. Finally, we describe some earlier work on the determination of
CFs and power spectra of galaxies, and discuss the problem of whether 3D CFs
can be calculated from 2D CFs.

\subsection{Spatial and projected density fields}

The essential feature of the cosmic web is its spatial structure or
pattern.  The spatial CF characterises the geometrical properties of the
cosmic web in a general and global way.  To understand the influence
of the pattern of the cosmic web on properties of CFs, let us
look at the geometry of the density field of the cosmic web, as given
by 2D and 3D data.

 Figure~\ref{fig:Fig9} shows 2D density fields of the model LCDM.10
using particle density limit $\rho_0=10$, and at various
thicknesses. The upper panels present 2D density fields of thin slices
of thickness $L=8$~\Mpc\ in $x,~y$ coordinates at three $z$ locations.
The thickness of the 2D density field in the bottom left panel is
$L=512$~\Mpc, that is, the whole cube of our LCDM.10 model.  In the bottom
middle and bottom right panels, the thicknesses are $L=128$~\Mpc\ and
$L=32$~\Mpc, respectively.  In the calculation of the 2D density
fields we used the 3D 
density field with a resolution of
$512 \times 512$; in the figure we show only the central
$256 \times 256$~\Mpc\ sections of the fields.  Galaxies can form in regions
where the local density is high enough; a particle density limit of
$\rho_0=10$ corresponds approximately to $L^\ast$ galaxies of SDSS magnitude
$M_r -5 \log h =-20.5$ using the percolation test; see Fig.~10 by
\citet{Einasto:2019aa}.  Thus, in Fig.~\ref{fig:Fig9} all DM particles
in low-density regions and DM particles corresponding to faint
galaxies, $M_r > -20.5$, are excluded.  In this way,
Fig.~\ref{fig:Fig9} imitates the 2D distribution of matter in real galaxies in
samples of different thickness.

Two-dimensional density fields in the top panels of Fig 3.4.4 are so thin that their morphological
properties are close to those of the 3D density field of the same
model, namely the fraction of high- and low-density regions and mutual distances
between high-density regions.  However, as seen from Table~\ref{Tab3},
the mean amplitude of CFs of stacked 2D LCDM.10.0064 model sheets of
thickness 8~\Mpc\ differs slightly from the amplitude of the 2D model
LCDM.10.2048, which is approximately equivalent to the 3D model
LCDM.10; compare Tables \ref{Tab1} and \ref{Tab3}.  This means that at
a thickness $L=8$~\Mpc, the\ 3D features of the web are already partly 
smoothed out in the 2D field.

Essential elements of the cosmic web are high-density regions
(clusters and filaments) and voids, which together form the pattern of
the web.  The detailed structure of the web is seen in the top panels of
Fig.~\ref{fig:Fig9}.  High-density regions are surrounded by
zero-density voids, occupying most of the volume of the density
field. 
{The fraction of cells with zero-density in 2D density fields of
  thickness 8~\Mpc\ is $F_0=0.80$; see Fig.~\ref{fig:Fig8}, and the filling
  factor of voids in the LCDM 3D density field is $F_V=0.86$; see
  Table 1 of \citet{Einasto:2018aa}.  } There are faint filaments of
DM in voids, but the density is too low to start galaxy formation, and therefore the
density field of galaxies has zero density there.  The characteristic
scale of halos containing galaxies is a few \Mpc; small visible
regions in Fig.~\ref{fig:Fig9} have approximately this diameter.
Clusters and filaments in sheets at different $z$ locations are
located in various $x,y$-positions; compare different top panels of
Fig.~\ref{fig:Fig9}.

Correlation functions are sensitive to particle/galaxy { separations}, not {
  locations}. For this reason, the statistical properties of the 2D CFs of very
thin sheets at various $z$-locations are similar, and close to
statistical properties of 3D correlation functions.  This statistical
similarity can be seen in the mean CF of $n=2048$ thin sheets of
thickness $L=512/n=0.25$~\Mpc, and is shown in Fig.~\ref{fig:Fig6}
together with the 3D CF of the same particle density limit $\rho_0$.

{With increasing thickness $L$ of 2D density fields, the fraction
  of cells with zero density rapidly decreases.  This is clearly seen
  in Fig.~\ref{fig:Fig9} visually, and in Fig.~\ref{fig:Fig8}
  graphically.   The fraction of zero-density cells is
  $F_0=0.53,~0.13,~5.4 \times 10^{-4}$ for thicknesses $L=32,~128,~512$~\Mpc\ of
  the 2D density field of the model LCDM.10. For models LCDM.05 and
  LCDM.02, there are  already no zero-density cells at thicknesses
  $L=512$ and $L=256$~\Mpc, respectively; see right panel of
  Fig.~\ref{fig:Fig8}. } 

\subsection{Correlation function amplitude as a cosmological parameter}

Amplitudes of CFs depend on many factors: (i) cosmological parameters:
matter-energy densities $\Omega_b,~\Omega_m,~\Omega_{\Lambda}$, and
the present {\em rms} matter fluctuation amplitude averaged over a
sphere of radius 8~\Mpc, $\sigma_8$; (ii) luminosities of galaxies;
(iii) systematic motions of galaxies in clusters -- the FOG effect,
and the flow of galaxies toward attractors -- the Kaiser effect.
Systematic motions decrease the amplitudes of 3D CFs on small scales
and increase them on large scales.  The increase of CF amplitudes on large
scales is well described by the dynamical model by
\citet{Kaiser:1987aa}.

The present analysis shows that amplitudes of 2D CFs are influenced by
an additional factor: the thickness of samples, $L$.  The dependence
of  the amplitudes of 2D CFs on sample thickness follows from the
spatial properties of the cosmic web. The cosmic web consists of
galaxies located in an intertwined filamentary pattern, and leaving
most of the space void of galaxies.  In projection, clusters and
filaments fill in voids, depending on the thickness of samples.  For
this reason, 2D and 3D patterns of the web are qualitatively different.
The thicker the 2D sheets, the greater the difference.

The filling factor of high-density regions of the model LCDM.05 is
0.10;  the SDSS sample  has a similar filling factor with absolute
magnitude limit $M_r=-19.0$; see \citet{Einasto:2019aa}.  The rest of
the volume, that is, 90\%, has zero density.  This means that 3D density
fields corresponding to galaxies as well as thin slices of the 2D
density field are dominated by zero-density cells.  In stacked thick 2D
sheets, clusters and filaments at various $z$ are projected to the 2D
$x,y$ plane at different positions, and in this way fill in voids in
the 2D density field.  This is clearly seen in the bottom panels of
Fig.~\ref{fig:Fig9}.  

The correlation functions of our models were calculated using density fields applying the
method by \citet{Szapudi:2005aa}.  The power spectrum of our LCDM
model was calculated by \citet{Einasto:2019aa}.  As input for both
correlation and power spectrum analysis, we use the density contrast field,
$\delta =(N-\bar{N})/\bar{N}$, used for power spectrum analysis as
follows:
\begin{equation}
P(k)=\langle |\delta_k|^2\rangle,
\end{equation}
where $k$ is the wavenumber, $\delta = \rho - 1$ is the density
contrast, and $\rho$ is the density in mean density units.  The
density field used to find power spectra or CFs can be divided into
four main regions: zero-density regions with $\rho=0$ and $\delta=-1$,
low-density regions with $0 \le \rho \le 2$ and $|\delta| \le 1$,
medium-density regions with $2 \le \rho \le 10$, and high-density
regions with $\rho > 10$.  Both the power spectrum and the CF depend
on fractions of different density regions.  In the full DM model, all
basic regions are present.  Over most of the volume, the density is
less than the mean density, or exceeds the mean density only slightly,
$\rho \le 2$.  In these low-density regions, density contrast lies in
the interval $-1 \le \delta \le 1$, and has a mean value of
$|\delta| \approx 0.5$ or less.  Here, matter remains in diffuse form,
galaxy formation is impossible, and the density field of galaxies has
zero density, $\rho =0$, $\delta=-1$, and $|\delta|=1$.  All cells of
the DM density field, which had density contrast in the interval
$-1 \ge \delta \ge 1$, have a value
$|\delta| =1$  in the galaxy density field.  The power spectrum (and CF) is a sum over all density
contrasts. The greater the fraction of cells with zero density, the
higher the amplitude of power spectra and CFs.  When we consider
density fields of increasing luminosity (particle density) limit, with
increasing luminosity limit, we see an increasing fraction of 
medium density cells, which  previously had densities in the interval
$2 \le \rho \le 10$, also a change in the number of zero-density cells with
$|\delta| =1$, which leads to a further increase in the amplitude of
the power spectra and CFs.

This analysis shows that  amplitudes of CFs and power spectra depend
essentially on the fraction of cells of the density field with zero density. 
This is the reason why the amplitudes of the CFs and power spectra
 of galaxies are higher than the amplitudes  of the CFs and power spectra
 of DM, and also why the amplitudes of CFs and power spectra of 
luminous galaxies are higher than the amplitudes of CFs and power
spectra  of fainter galaxies.

The same argument is valid in the comparison of 2D and 3D density fields. 
The essential difference between 2D and 3D density fields is in the
fraction of zero-density regions.  In 2D density fields, high-density
regions from various distances overlap and fill in voids. The thicker
the 2D density fields, the fewer zero-density cells  they contain. We
conclude that the essential difference between 2D and 3D (and thin 2D)
density fields is the near absence of visible zero-density regions
in thick 2D fields.  To summarise the comparison of 2D and 3D density
fields, we can say that the large fraction of zero-density regions of
3D fields, wiped out in various amounts in 2D fields, decreases the
amplitude of 2D CFs; this effect increases with increasing
thickness of the 2D field.  For this reason, the { amplitudes of power
  spectra and CFs for stacked 2D density fields are lower than
  the amplitudes of power spectra and 3D CFs based on 3D density fields. }
{ For the same reason,   the amplitudes of the power spectra and
  3D  CFs of galaxies are greater than the amplitudes of  3D CFs of DM.
 }

The similarity between the 2D CFs of Millennium simulation galaxies and the 2D CFs
of LCDM samples shows that the influence of zero-density regions in
both models is similar, also for Millennium models in redshift space.
An analytical description of the relation between 2D and 3D two-point
correlators is given in the Appendix.

\subsection{Comparison with earlier work}

{In classical correlation studies, the first step was to find the
  estimate of the anisotropic 3D CF by counting pairs $DD(r_p, \pi)$ 
  \citep{Davis:1983ly}. The next step was the calculation of the 2D CF
  using Eq.~(\ref{eq12}).  As a final step \citet{Davis:1983ly}
  calculated the 3D CF using the inversion according to Eq.~(\ref{eq13}),
  with  smoothed $w(r_\perp)$ 2D CF.
}

\citet{Norberg:2001aa} selected galaxies in conical shells of various
thicknesses from $r_{max}- r_{min}\approx 25$~\Mpc\ for the faintest
galaxies ($M_b - 5\log h \approx -18$) to
$r_{max}- r_{min}\approx 700$~\Mpc\ for the most luminous subsamples.
\citet{Zehavi:2005aa, Zehavi:2011aa} investigated projected CFs of
SDSS galaxies of different luminosity.  Authors applied standard 
practice by \citet{Davis:1983ly} and computed projected CFs using Eq.~(\ref{eq12}), and
real-space correlation functions using Eq.~(\ref{eq13}).  Samples of
various absolute magnitude bins were located in spherical shells of
various thicknesses.

A correlation analysis of SDSS samples using 3D CFs in redshift space
was performed by \citet{Hawkins:2003aa} and \citet{Einasto:2020aa}.  These
analyses showed that correlation radii from 3D analysis in redshift
space are about a factor 1.35 higher than from the 2D analysis by
\citet{Zehavi:2011aa}.  This difference is probably due to projection
effects in the 2D analysis.

\citet{Shi:2016aa,Shi:2018aa} elaborated a method to calculate
real-space two-point CFs of galaxies from redshift data. The method
consists of several steps: calculating the mass-density field from the
luminosity density field, reconstructing the velocity field on
quasi-linear scale, and correction for the Kaiser and FOG effects. The method
was checked using mock galaxy catalogues based on a simulation of
$3072^3$ particles in a box of 500~\Mpc\ side-length.  Two-point
spatial CFs of mock catalogues in true real space, redshift space,
Kaiser space, and FOG space are relatively similar to the corresponding CFs in
our study, shown in Figs.~\ref{fig:Fig1} and \ref{fig:Fig4}. 

\subsection{The information content of 2D CFs}

{Our previous analysis showed that calculation of projected CFs
  from spatial CFs influences the information content of CFs. We   now
  try to estimate the change of information content in all steps of
  the 
  standard procedures by \citet{Davis:1983ly}.  The anisotropic CF,
  $\xithree(\parl{r},r_\perp)$, found by counting pairs of separations
  of galaxies, has the maximal information content possible from
  observational data in terms of angular positions and redshifts of
  galaxies.  Observational data are influenced by redshift
  distortions.  To avoid distortions, the $\xithree(\parl{r},r_\perp)$
  is integrated along the line of sight (see Eq.~(\ref{2dens}) or
  Eq.~(\ref{eq12})), to get 2D CFs. 
  Our analysis shows that this operation preserves information on
  the luminosity dependence of the CF, which is confirmed by comparison of
  correlation radii of galaxies of various luminosity obtained by
  \citet{Davis:1983ly}, \citet{Norberg:2001aa}, and
  \citet{Zehavi:2005aa, Zehavi:2011aa}, with the results of authors using
  different methods \citep{Hawkins:2003aa}.

  Our study also shows that integration according to Eq.~(\ref{eq04})
  yields 2D CFs with decreasing amplitudes, depending on the thickness
  of the sheets.  This means that, because of projection effects, information
  on zero-density regions in the 3D density field is partly lost, as
  demonstrated graphically in Fig.~\ref{fig:Fig9}, and numerically in
  Figs.~\ref{fig:Fig6} and \ref{fig:Fig8}.  In real observational
  samples, spherical coordinates are used and integration is done
  according to Eq.~(\ref{eq12}).  The projection effect to 2D CFs is
  smaller than expected from Table~\ref{Tab4}, because observed galaxy
  samples are conical, and nearby regions in conical shells have a
  smaller input than more distant ones.  Thus, the decrease in  the
  amplitude of 2D CFs is not well determined and depends on 
  the sample used.  }

{We also find that in 2D density fields, the information
  on the internal structure of clusters of galaxies is lost,
  especially in redshifts space; see the behaviour of gradient
  functions in Fig.~\ref{fig:Fig6}.  This explains the observational
  result by \citet{Maddox:1990aa} that the angular CF of the
  Automatic Plate Measuring (APM) survey of galaxies has a constant
  slope, $\gamma \approx -0.7$, over a wide range of angular scales
  and galaxy apparent magnitudes.  As information on the internal
  structure of clusters is lost, this result can be explained as
  evidence for a mean constant fractal dimension of the spatial
  distribution of galaxies of the survey.}
 
{Three-dimensional density fields contain more information than 2D density
  fields, and therefore inversion to get 3D CFs from 2D CFs (Eq.~(\ref{eq13}))
  adds information.  The relation between projected and spatial CFs is
  similar to the relation between the projected and spatial densities of
  galaxies, which are used to calculate their dynamical models, starting from
  \citet{Wyse:1942aa} and \citet{Kuzmin:1952aa}.  The information gain
  comes in this case from the assumption that galaxies are spatial
  ellipsoids of rotation, which was specifically noted by
  \citet{Kuzmin:1952aa}.  The information gain in the calculation of 3D CFs
  from 2D CFs using Eq.~(\ref{eq13}) comes from the tacit assumption
  that the spatial and projected structures of the cosmic web are
  statistically similar, which is not the case.  The difference between 2D and 3D data can be
  treated in information terms: information on zero-density regions is
  lost in 2D density fields and in respective 2D CFs. This information
  loss is characterised by decreasing amplitudes of 2D CFs.  When 2D CFs
  are used to calculate 3D CFs, this information on 3D density fields
  is not restored, and therefore amplitudes of 3D CFs found by inversion
(see  Eq.~(\ref{eq13})) are still based on amplitudes of 2D CFs distorted
  by projection effects. In other words,  it is impossible to
  calculate 3D density fields, which are
close to sheets in the top panels
  of Fig.~\ref{fig:Fig9} using 2D density fields, as
  shown in the lower panels. }

\section{Conclusions}

In this paper, we analyse the relationship between projected and
spatial CFs.  To avoid complications related to observational
selection effects we studied the relationship using simulated DM
models, using  our own LCDM model and the Millennium simulation.
Both simulations were made in a box of size $\sim 500$~\Mpc. To study
spatial distributions of model objects we used DM particles in our
model and simulated galaxies in the Millennium model.

{We calculated CFs with the \citet{Szapudi:2005aa} method, which
  allows us to consider CFs as descriptors of the density field of the
  cosmic web.}  We calculated 3D CFs for both models, for the
Millennium model in real and redshift space. We found projected 2D CFs
in the distant observer approximation in the $z$-direction.  In
addition to CFs, we used gradient functions,
$\gamma(r)= = {\dd{\log g(r)} / \dd{\log r}}$, where $g(r)=1+\xi(r)$,
and relative bias functions,
$b_R(r, \rho_0) = \sqrt{\xi_C(r,\rho_0)/\xi_r(r)}$, where
$\xi_C(r,\rho_0)$ is the CF of clustered matter (galaxies), and
$\xi_r(r)$ is the CF of the reference sample.  Additionally, we
quantify CFs by their amplitudes at separation 6~\Mpc, $A=\xi(6)$.

The general results of our study can be summarised as follows.

\begin{enumerate}

\item{} The dominant elements of the cosmic web are clusters and
  filaments, which are separated by voids filling most of the
  volume. Clusters and filaments are located at different 
  positions in 
   2D sheets.  As a result, in projection, clusters and filaments fill in
  2D voids, which leads to a decrease in the amplitudes of the CFs
  (and power spectra).  For this reason, the amplitudes of 2D CFs 
  are lower than the amplitudes of 3D CFs, and the
thicker the 2D sample, the greater the  difference.

\item{} {In 2D CFs, information on the luminosity dependence is
    preserved, information on the internal structure of clusters is
    lost, and information on the amplitudes of CFs is partially lost,
    depending on the thickness of the sample.}

\end{enumerate}

{The amplitude of CFs (and power spectra) is an important
  cosmological parameter which depends on many factors -- cosmological
  parameters, systematic motions of galaxies, luminosities of
  galaxies, and thickness of 2D samples.  Luminosities and thickness
  influence 2D CFs in oposite directions: increasing luminosity
  increases, but  increasing thickness decreases CF amplitudes.
}

\begin{acknowledgements} 

Our special thanks are to Enn Saar for many  discussions,
and to  anonymous referees for stimulating suggestions that
greatly improved the  paper. We thank the Millennium team for making
results of their simulations public.

This work was supported by institutional research funding IUT26-2 and
IUT40-2 of the Estonian Ministry of Education and Research, and by the
Estonian Research Council grants PRG803, PRG1006 and MOBTT5.  We acknowledge
the support by the Centre of Excellence ``Dark side of the Universe''
(TK133) financed by the European Union through the European Regional
Development Fund.

\end{acknowledgements} 
 
\bibliographystyle{aa} 

\nopagebreak
\onecolumn
\begin{appendix}
 
  \section{Two-point function of the projected field}
  
  Here we briefly describe the relation between 2D and 3D two-point
  correlators. In the following we assume a sufficiently small survey
  volume such that evolutionary and lightcone effects can be
  neglected. We assume the validity of the {\emph cosmological
    principle}, that is, statistical homogeneity and isotropy of the
  cosmic density 
  field. Additionally, we do not include redshift-space distortions,
  because we focus on real-space two-point functions.

The density contrast is defined as is customary:
\be
\dthree({\bf r})\equiv\frac{n({\bf r})}{\bar{n}}-1\,,
\ee
where $n({\bf r})$ is the comoving number density of tracer objects at
spatial location ${\bf r}$ and $\bar{n}$ is the corresponding average
density. 

In the following we consider plane-parallel and spherical projections
of the 3D field separately.

\subsection{Plane-parallel geometry}

Assuming appropriately normalized projection and selection function $w(r)$, namely
\be
\int\dd{r} w(r)\equiv 1\,,
\ee
the projected 2D overdensity field can be expressed as
\be
\dtwo(\per{r})=\int\dd{\parl{r}}w(\parl{r})\dthree(\parl{r},\per{r})\,.
\ee

The two-point correlation function of the projected density field can
now be obtained. According to the cosmological principle we can always
choose one point to be at the origin, that is, ${\per{r}}_{,1}={\bf 0}$,
and for the other we assume ${\per{r}}_{,2}={\bf R}$ with modulus
$R\equiv |{\bf R}|$. Thus, one can write 
\bea\label{eq04}
\xitwo(R)&=\langle\dtwo({\bf 0})\dtwo({\bf R})\rangle=\int\dd{r_1}w(r_1)\int\dd{r_2}w(r_2)\langle\dthree(r_1,{\bf 0})\dthree(r_2,{\bf R})\rangle=\\
&=\int\dd{r_1}w(r_1)\int\dd{r_2}w(r_2)\xithree\left(\sqrt{R^2+(r_2-r_1)^2}\right).
\eea
In the case of a uniform selection within the range $[0,L]$, that is,~$w(r)= 1/L$, the above result can be expressed as
\be\label{eq05}
\xitwo(R)=\int_0^1\dd{x_1}\int_0^1\dd{x_2}\xithree\left(\sqrt{R^2+L^2(x_2-x_1)^2}\right)\,.
\ee
In Eq.~(\ref{eq04}) the $\xithree$ part of the integrand is peaked
around $r_1=r_2$. If the selection function varies smoothly in
comparison, it can be pulled out of the second integral (this is the
essence of the Limber approximation, \eg~\citet{Peebles:1980aa}),
giving 
\be
\xitwo(R)\simeq\int_0^\infty\dd{r}w^2(r)\int_{-\infty}^\infty\dd{x}\xithree\left(\sqrt{R^2+x^2}\right)=2\int_0^\infty\dd{r}w^2(r)\int_0^\infty\dd{x}\xithree\left(\sqrt{R^2+x^2}\right)\,, 
\ee
where we have introduced new dummy variables $r_1\rightarrow r$ and $r_2-r_1\rightarrow x$. In the case of a uniform selection within the range $[0,L],$ the above result can be expressed as
\be\label{eq07}
\xitwo(R)\simeq\frac{2}{L}\int_0^\infty\dd{x}\xithree\left(\sqrt{R^2+x^2}\right)=\frac{2}{L}\int_R^\infty\dd{r}\frac{r}{\sqrt{r^2-R^2}}\,\xithree(r)
.\ee

\subsection{Spherical geometry}

The above results can be extended for the case of spherical
geometry. Here the results have the simplest form once the radial
selection and projection function is normalized as 
\be
\int\dd{r}r^2w(r)\equiv 1\,.
\ee
In that case the analogue of Eq.~(\ref{eq04}) reads
\be\label{eq09}
\xitwo(\Theta)=\int\dd{r_1}r_1^2w(r_1)\int\dd{r_2}r_2^2w(r_2)\xithree\left(\sqrt{r_1^2+r_2^2-2r_1r_2\cos{\Theta}}\right)\,,
\ee
where $\cos{\Theta}\equiv {\bf\hat{r}_1}\cdot{\bf\hat{r}_2}$ with
${\bf\hat{r}_1}$,${\bf\hat{r}_2}$ the radial unit vectors marking the
two points. 

The case with uniform selection in the range $[D-L/2,D+L/2]$,
that is, the~analogue of Eq.~(\ref{eq05}), can be written as
\be
\xitwo(R)=\int_{D-L/2}^{D+L/2}\dd{r_1}r_1^2\int_{D-L/2}^{D+L/2}\dd{r_2}r_2^2\,\xithree\left(\sqrt{r_1^2+r_2^2-2r_1r_2\cos\left(\Theta={R/D}\right)}\right)\,.
\ee
In the case of slowly varying selection and with small-angle approximation, Eq.~(\ref{eq09}) can be recast as follows (this is a standard
Limber's formula): 
\be
\xitwo(\Theta)\simeq\int_0^\infty\dd{r}r^4w^2(r)\int_{-\infty}^\infty\dd{x}\xithree\left(\sqrt{x^2+r^2\Theta^2}\right)\,,
\ee
where new dummy variables $r_1\rightarrow r$ and $r_2-r_1\rightarrow x$ were introduced.

\section{Projected two-point function}

Very often, in order to avoid complications caused by
redshift-space distortions, the spatial two-point CF
which is evaluated as a 2D function of radial and transverse
separations, is integrated along the radial direction, resulting in
the following quantity: 
\be\label{eq12}
w(r_\perp)=\int_{-\infty}^\infty\xithree(\parl{r},r_\perp)\,\dd{\parl{r}}=2\int_{r_\perp}^\infty\dd{r}\frac{r}{\sqrt{r^2-r_\perp^2}}\,\xithree(r)\,.
\ee
The above has a form of the Abel integral equation, which is often
inverted to recover the real-space correlation function. The analytic
form for the inversion is given by the following relation 
\be
\label{eq13}
\xithree(r)=-\frac{1}{\pi}\int_r^\infty\dd{r_\perp}\frac{1}{\sqrt{r_\perp^2-r^2}}\frac{\dd{w(r_\perp)}}{\dd{r_\perp}}\,. 
\ee
However, in the presence of measurement errors this inversion is not a
well posed problem (note that it involves taking derivatives from the
noisy data), and thus needs additional assumptions for regularisation
purposes (often one assumes a specific functional form for $\xithree$,
for example a~simple power law). In addition, instead of using all available
information, which would demand somewhat more complicated modelling in
order to also capture the signal stored in the redshift-space distortions,
the line-of-sight integration results in a generic signal loss. 

It is instructive to note that the above result~(\ref{eq12}) is
almost identical to Eq.~(\ref{eq07}). This is only so under the
validity of the Limber approximation, that is, in the case where the projection
and correlator calculation operations effectively commute. In general,
there is a clear difference between the projected CF
and the CF of the projected field.

\end{appendix}
\end{document}